\documentclass[aps,prc,superscriptaddress,twoside,twocolumn,nofootinbib,10pt,showpacs,floatfix]{revtex4-1}

\usepackage{braket}
\usepackage{bm}
\usepackage{amsmath}
\usepackage[pdftex]{graphicx}
\usepackage{multirow}
\newcommand{\Slash}[1]{{\ooalign{\hfil/\hfil\crcr$#1$}}}
\usepackage{colortbl}
\usepackage{here}
\usepackage{url}
\usepackage{ulem} 
\usepackage{amssymb}
\allowdisplaybreaks

\begin{document}

	\title{Constraint to chiral invariant masses of nucleons from GW170817 in an extended parity doublet model}

	\author{Takahiro Yamazaki}
	\email{yamazaki@hken.phys.nagoya-u.ac.jp}
	\author{Masayasu Harada}
	\email{harada@hken.phys.nagoya-u.ac.jp}
	\affiliation{Department of Physics,  Nagoya University, Nagoya, 464-8602, Japan}
	\date{\today}

	\newcommand\sect[1]{\emph{#1}---}
\begin{abstract}
We construct nuclear matter based on an extended parity doublet model including four light nucleons $N(939)$, $N(1440)$, $N(1535)$, and $N(1650)$. 
We exclude some values of the chiral invariant masses by requiring the saturation properties
of normal nuclear matter; 
saturation density, binding energy, incompressibility, and symmetry energy.
We find further constraint 
to the chiral invariant masses from the  tidal deformability determined by the observation of the gravitational waves from neutron star merger GW170817.
Our result shows that the chiral invariant masses are larger than about $600\,$MeV.
We also give some predictions on the symmetry energy and the slope parameters in the high density region, which will be measured in future experiments.
\end{abstract}
\maketitle
\section{Introduction}
\label{sec:Intro}

The chiral symmetry and its spontaneous breaking is one of the important features in low-energy hadron physics based on QCD.
The breaking generates a part of hadron masses and causes a splitting between chiral partners.
It is interesting to ask how much of the nucleon mass is generated by the spontaneous chiral symmetry breaking and what is the chiral partner to the nucleon.

In Ref.~\cite{Detar:1988kn}, a model based on the parity doublet structure was introduced, where the excited nucleon $N(1535)$ is regarded as the chiral partner to the nucleon $N(939)$. It is important to note that their masses include a chiral invariant mass in addition to the  masses caused by the spontaneous chiral symmetry breaking.  
The determination of the chiral invariant mass using the phenomenology at vacuum is done in e.g. Refs.~\cite{Jido:1998av,Jido:2001nt,Gallas:2009qp},
which shows that the chiral invariant mass of the nucleon is smaller than about $500$\,MeV.

The parity doublet structure is extended to include hyperons and/or more nucleons 
in e.g., Refs.~\cite{Nemoto:1998um,Chen:2008qv,Dmitrasinovic:2009vp,Dmitrasinovic:2009vy,Chen:2009sf,Chen:2010ba,Steinheimer:2011ea,%
Chen:2011rh,Nishihara:2015fka,Olbrich:2015gln,Dmitrasinovic:2016hup,Sasaki:2017glk,Yamazaki:2018stk}.
In Ref.~\cite{Yamazaki:2018stk}, the authors of present paper constructed a model which includes two chiral representations, 
the $[(\bf{2},\bf{3})\oplus(\bf{3},\bf{2})]$ representation under $\mbox{SU}(2)_L \otimes \mbox{SU}(2)_R$ in addition to $[(\bf{1},\bf{2})\oplus(\bf{2},\bf{1})]$ representation, to study four nucleons, $N(939)$, $N(1440)$, $N(1535)$ and $N(1650)$.
It was shown that there are wide range of two chiral invariant masses satisfying vacuum properties of the nucleons; the masses, the axial charges and the pionic decay widths, and that the solutions are categorized into five groups.

The properties of hot and/or dense matter including neutron star matter 
based on the parity doublet structure 
are widely studied in 
Refs.~\cite{Hatsuda:1988mv, Zschiesche:2006zj, Dexheimer:2007tn, Dexheimer:2008cv, Sasaki:2010bp, Sasaki:2011ff,%
Gallas:2011qp, Paeng:2011hy,%
Steinheimer:2011ea,Dexheimer:2012eu, Paeng:2013xya,Benic:2015pia,Motohiro:2015taa,%
Mukherjee:2016nhb,Suenaga:2017wbb,Takeda:2017mrm,Mukherjee:2017jzi,Marczenko:2017huu,Abuki:2018ijb,Marczenko:2018jui}.
In ~\cite{Hatsuda:1988mv, Zschiesche:2006zj, Dexheimer:2007tn, Dexheimer:2008cv, Sasaki:2010bp, Gallas:2011qp,Steinheimer:2011ea, Dexheimer:2012eu, Benic:2015pia}, 
the authors 
studied the relation between the chiral invariant mass and the incompressibility $K$ of nuclear matter and their results show that  the empirical value $K \sim 240$\,MeV only when the chiral invariant mass is close to nucleon's mass, $m_0\sim900$\,MeV.
In  Ref.~\cite{Motohiro:2015taa},
six point interaction of scalar mesons was introduced and it was shown that the saturation properties are reproduced for wide region of the chiral invariant mass, i.e. $500 \le m_0 \le 900$\,MeV.
In Ref.~\cite{Marczenko:2018jui}, the constraint to the chiral invariant mass from the properties of neutron stars including the tidal deformability observed from 
GW170817~\cite{TheLIGOScientific:2017qsa,GBM:2017lvd,Abbott:2018exr} was obtained as
$780$-$810$\,MeV.
Recently, the parity doublet model is used to study the nuclei with finite size in Ref.~\cite{Shin:2018axs} which shows that $m_0 \sim 700$\,MeV is preferred to reproduce the properties of nuclei.
The relatively large value of the chiral invariant mass seems also consistent with lattice analyses in Refs.~\cite{Aarts:2017rrl,Aarts:2017iai,Aarts:2018glk}.

In this paper, we construct nuclear matter and neutron star matter using the model introduced in Ref.~\cite{Yamazaki:2018stk} based on the mean field approximation.
For the meson parts, we use the model introduced in Ref.~\cite{Motohiro:2015taa}: the six point interaction of the scalar field is introduced and the $\omega$ and $\rho$ mesons are included based on the hidden local symmetry~\cite{Bando:1987br,Harada:2003jx}.
We will show that requiring the saturation properties of normal nuclear matter excludes some combinations of two chiral invariant masses.
We solve the Tolman-Oppenheimer-Volkov (TOV) 
equation~\cite{Tolman:1939jz,Oppenheimer:1939ne} to determine the energy density and tidal deformability of neutron stars.
Then, we will show that the tidal deformability observed from 
GW170817~\cite{TheLIGOScientific:2017qsa,GBM:2017lvd,Abbott:2018exr} provides further constraint to the chiral invariant masses.

This paper is organized as follows: 
In section~\ref{sec: model}, we include the $\omega$ and $\rho$ mesons into the model introduced in Ref.~\cite{Yamazaki:2018stk}.
We give formulations to study nuclear matter in the mean field approximation in section~\ref{sec: formulation}.
Here we show formulas to study saturation properties of the normal nuclear matter and the equation of state for neutron star matter.
Section~\ref{sec: numerical} is devoted to the main part where we obtain constraints to the chiral invariant masses from the saturation properties and tidal deformability from GW170817.
We also provide predictions for the relations between the mass and radius as well as those for the mass and the central density of neutron stars, and the symmetry energy and the slope parameter in dense matter.
Finally, we will give a summary and discussions in section~\ref{sec: summary}.

\section{model}
\label{sec: model}

In this section, we introduce
an extended parity doublet model to describe nuclear matter based on the model constructed in 
Ref.~\cite{Yamazaki:2018stk}.
The model includes four baryon fields corresponding to the following representations under SU(2)$_L\times$SU(2)$_R$ chiral symmetry:
\begin{align}
	\psi_{1l}\sim({\bf 2},{\bf 1}),\hspace{10mm}\psi_{1r}\sim({\bf 1},{\bf 2}) \ ,
	\nonumber
	\\
	\psi_{2l}\sim({\bf 1},{\bf 2}),\hspace{10mm}\psi_{2r}\sim({\bf 2},{\bf 1}) \ , 
	\nonumber
	\\
	\eta_{1l}\sim({\bf 2},{\bf 3}),\hspace{10mm}\eta_{1r}\sim({\bf 3},{\bf 2}) \ , 
	\nonumber
	\\
	\eta_{2l}\sim({\bf 3},{\bf 2}),\hspace{10mm}\eta_{2r}\sim({\bf 2},{\bf 3}) \ .
	\label{assignment}
\end{align}
The iso-singlet scalar meson $\sigma$ and the iso-triplet pseudoscalar meson $\pi$ are included in a matrix field $M$, which transforms as
\begin{equation}
M \to g_L M g_R^\dag \ ,
\end{equation}
where $g_{L,R} \in \mbox{SU(2)}_{L,R}$.
Following Ref.~\cite{Motohiro:2015taa}, we include $\omega$ and $\rho$ mesons as the gauge bosons of hidden local symmetry~\cite{Bando:1987br,Harada:2003jx} by performing the polar decomposition of the field $M$ as~\footnote{
The normalization of the $M$ field in this paper is a half of the one in Ref.~\cite{Motohiro:2015taa}.
}
\begin{align}
	M = \xi_L^{\dag}\frac{\sigma}{2}\xi_R = \frac{\sigma}{2} \xi_L^\dag \, \xi_R = \frac{\sigma}{2} \, U \ .
\end{align}
We introduce the same potential for $M$ as used in Ref.~\cite{Motohiro:2015taa}: 
\begin{align}
V_M = & -\bar{\mu}^2{\rm tr}[MM^{\dag}]+\lambda_4[{\rm tr}[MM^{\dag}]]^2-\frac{4}{3}\lambda_6[{\rm tr}[MM^{\dag}]]^3 \notag\\
& {} -\frac{1}{2}\epsilon({\rm tr}[\mathcal{M}^{\dag}M]+{\rm tr}[\mathcal{M}M^{\dag}])
\ , 
\end{align}
where $\epsilon$ is a parameter with dimension two and ${\mathcal M}$ is the quark mass matrix defined as 
\begin{align}
	\mathcal{M}=
		\begin{pmatrix}
		m_u	&&	0
		\\
		0	&&	m_d
		\end{pmatrix}
\end{align}
with $m_u$ and $m_d$ begin the masses of up and down quarks.
In the present analysis, we neglect the difference between these masses, and take $m_u = m_d = \bar{m} $. 
In the vacuum, the combination $\bar{m} \epsilon$ is related the pion mass as
\begin{equation}
\bar{m} \epsilon = m_\pi^2 f_\pi \ .
\end{equation} 
We adopt the Yukawa interaction terms among $M$ and the nucleons as in Ref.~\cite{Yamazaki:2018stk}, so that we omit those in this paper.

We introduce the interaction terms among the vector mesons and nucleons similarly to 
Ref.~\cite{Motohiro:2015taa}.
Here, instead of writing full Lagrangian, we shall show the relevant terms in the present analysis.
The resultant interaction terms for $\omega$ meson is written as
\begin{align}
	\mathcal{L}_{\omega N}=-g_{\omega}(\sum_{i=1,2}\bar{\psi_i}\Slash{\omega}\psi_i+\sum_{j=1,2}\bar{\eta_j}\Slash{\omega}\eta_j) \ .
\end{align}
Here we assume that the coupling to $\psi$ is the same as that to $\eta$ for simplicity.
Similarly, the interactions for $\rho$ mesons is given by
\begin{align}
	\mathcal{L}_{\rho N}=-\frac{1}{2}g_{\rho}(\sum_{i=1,2}\bar{\psi_i}\bm{\tau}\cdot\Slash{\bm{\rho}}\psi_i
	+\sum_{j=1,2}\bar{\eta_j}\bm{\tau}\cdot\Slash{\bm{\rho}}\eta_j) \ .
\end{align}
We note that the mass terms for $\omega$  and $\rho$ mesons are written as
\begin{align}
V_{\omega} = 
-\frac{1}{2}m_{\omega}^2\omega_{\mu}\omega^{\mu} \ , \quad
V_{\rho}=
-\frac{1}{2}m_{\rho}^2\bm{\rho}_{\mu}\bm{\rho}^{\mu} \ .
\end{align}

\section{Formulation}
\label{sec: formulation}

In this section, we present formulations to study nuclear matter in the mean field approximation based on the model introduced in the previous section.
Here we assume that all the parameters of the model do not depend on the chemical potentials.

\subsection{Thermodynamic Potential}

In the present analysis, 
we assume that there are no neutral and charged pion condensation, and that
the following fields have their vacuum expectation values (VEVs) as
\begin{align}
\sigma=\sigma_0 \, \quad \omega_{\mu =0} =\omega\ , \quad \rho_{\mu=0}^3=\rho \ .
\end{align}
In the mean field approximation, 
the thermodynamic potential is obtained by
\begin{align}
	\Omega=\sum_{i=1,2,3,4, \, N=p,n} \Omega_{N^{(i)}}+V_{M}+V_{\omega}+V_{\rho} \ , 
\end{align}
where
\begin{align}
V_M=& -\frac{\bar{\mu}^2}{2}\sigma_0^2+\frac{\lambda_4}{4}\sigma_0^4-\frac{\lambda_6}{6}\sigma_0^6 - m_{\pi}^2f_{\pi}\sigma_0 \ , 
\\
V_{\omega} = & 
-\frac{1}{2}m_{\omega}^2 \, \omega^2 \ ,
\\
V_{\rho}= &
-\frac{1}{2}m_{\rho}^2 \rho^2 \ .
\end{align}
The contribution from the nucleons, $\Omega_{N^{(i)}}$ is expressed as
\begin{align}
\Omega_{N^{(i)}}= 2\int\frac{d^3k}{(2\pi)^3}
\left( E_N^{(i)} - \bar{\mu}_N^{(i)} \right) \, \theta \left( \bar{\mu}_N^{(i)} - E_N^{(i)} \right) \ ,
\end{align}
where $\theta( x)$ is the step function defined as 
\begin{equation}
\theta (x) = \left\{ \begin{array}{ll} 1 & (x>0) \\ 0 & (x<0) \end{array} \right. \ , 
\end{equation}
$E_N^{(i)}$ is an energy of the nucleon
\begin{equation}
E_{N}^{(i)} = \sqrt{ k ^2 + \left(m_N^{(i)}\right)^2} \ .
\end{equation}
$\bar{\mu}_N^{(i)}$ is the effective chemical potential defined by
\begin{align}
\bar{\mu}_p^{(i)} = & \bar{\mu}_B + \frac{1}{2} \bar{\mu}_I \ , \notag\\
\bar{\mu}_n^{(i)} = & \bar{\mu}_B - \frac{1}{2} \bar{\mu}_I \ , 
\end{align}
with
\begin{align}
\bar{\mu}_B & = \mu_B - g_\omega \omega \ , \notag\\
\bar{\mu}_I & = \mu_I - g_\rho \rho \ .
\end{align}

We should note that, in the above expression, the mean fields $\sigma_0$, $\omega$ and $\rho$ are solutions of the stationary conditions:
\begin{align}
0=& \frac{\partial \Omega}{\partial \sigma_0} =  -\bar{\mu}^2\sigma_0 + \lambda_4\sigma_0^3-\lambda_6\sigma_0^5-m_{\pi}^2f_{\pi} \notag\\
& \ {} + 2 \sum_{i,N}\frac{\partial m_{N}^{{(i)}}}{\partial \sigma_0 } \int \frac{d^3k}{ (2\pi)^3 }  \frac{ m_N^{(i)} }{ E_N^{(i)} }  \, \theta \left( \bar{\mu}_N^{(i)} - E_N^{(i)} \right) \ , \\
0=& \frac{\partial \Omega}{\partial \omega} = -m^2_{\omega}\omega+g_{\omega} \rho_B  \ , \\
0=& \frac{\partial \Omega}{\partial \rho} = -m^2_{\rho}\rho+g_{\rho} \rho_I \ , 
\end{align}
where
\begin{align}
\rho_B = \sum_{i} \left( \rho_p^{(i)} + \rho_n^{(i)} \right) \ , \quad
\rho_I = \sum_{i} \frac{ \rho_p^{(i)} - \rho_n^{(i)} }{2} \ ,
\label{def:muBI}
\end{align}
with
\begin{align}
\rho_N^{(i)} = 2 \int\frac{d^3k}{(2\pi)^3} \, \theta \left( \bar{\mu}_N^{(i)} - E_N^{(i)} \right) \ .
\label{each density}
\end{align}

\subsection{Saturation Properties at Normal Nuclear Density}

In this subsection, we provide formulas to calculate several physical quantities of nuclear matter at normal nuclear density.

From the thermodynamic potential obtained in the previous section, the baryon number density and the isospin density are calculated as
\begin{equation}
\rho_B = - \left(  \frac{ \partial \Omega }{\partial \mu_B}  \right)_{\mu_I} \ , \quad
\rho_I = - \left(  \frac{ \partial \Omega }{\partial \mu_I}  \right)_{\mu_B} \ ,
\label{densityB}
\end{equation}
where $\left( \ \right)_{\mu_I}$ implies that the derivative in terms of $\mu_B$ is taken with fixed $\mu_I$, and similarly for $\left( \ \right)_{\mu_B}$.
One can easily confirm that $\mu_B$  and $\mu_I$ in Eq.~(\ref{densityB}) agree with those in Eq.~(\ref{def:muBI}).
The saturation density $\rho_0$ is calculated as
\begin{equation}
\rho_0 = \rho_B \left( \mu_B = \mu_0 \,,\, \mu_I = 0 \right) \ ,
\label{def rho0}
\end{equation}
where $\mu_0$ is the value of the baryon number chemical potential at saturation point.

The pressure of the system is given by
\begin{equation}
P = - \Omega \ .
\end{equation}
From the thermodynamic relation, the energy density  is obtained as
\begin{align}
	\epsilon=- P +\mu_B\rho_B+\mu_I\rho_I \ .
\end{align}
Then, the binding energy is given by
\begin{align}
E_{{\rm bind,}\rho_0} =\frac{E}{A} \bigg\vert_{\mu_B=\mu_0}-m_N
=\frac{\epsilon}{\rho_B} \bigg\vert_{\mu_B=\mu_0}-m_N \ .
\end{align}
From this, $\mu_0$ in Eq.~(\ref{def rho0}) is given as
\begin{equation}
\mu_0 = m_N^{(1)} - E_{{\rm bind},\rho_0} \ ,
\end{equation}
where $m_N^{(1)}$ is the mass of lightest nucleon.
We note that using above conditions, we can easily shows that the pressure at normal nuclear density vanishes:
\begin{equation}
P(\mu_B = \mu_0 \,,\, \mu_I = 0 ) = 0 \ .
\end{equation}
The incompressibility is calculated as
\begin{align}
K=9\rho_B^2\frac{\partial^2(\epsilon /\rho_B)}{\partial \rho_B^2} \bigg\vert_{\rho_B=\rho_0}=9\rho_B\frac{\partial \mu_B}{\partial \rho_B} \bigg\vert_{\rho_B=\rho_0}
\end{align}

The symmetry energy per nucleon is given as 
\begin{align}
E_{\rm sym}=& \frac{1}{2}\frac{\partial^2(\epsilon /\rho_B)}{\partial \delta^2} \bigg\vert_{\delta = 0}
=\frac{\rho_B}{8}\frac{\partial \mu_I}{\partial \rho_I} \bigg\vert_{\rho_I=0} \ ,
\end{align}
where $\delta$ is asymmetric parameter define as
\begin{align}
\delta\equiv\frac{\rho_p-\rho_n}{\rho_B}=\frac{2\rho_I}{\rho_B} \ .
\end{align}
From Eq.~(\ref{def:muBI}), this is calculated as
\begin{align}
E_{\rm sym}
=& \frac{\rho_B}{8}\left( \frac{2\pi^2}{\sum_{N , i}  k_{FN}^{{(i)}}E_{FN}^{{(i)}}}+\frac{g_{\rho}^2}{m_{\rho}^2} \right)\ , 
\label{eq:Esym}
\end{align}
where the summation is taken over $N = p , n$ and $i = 1,2,3,4$, and 
$k_{FN}^{(i)} $ and $E_{FN}^{(i)}$ are the Fermi momentum and Fermi energy of the nucleon.
The slope parameter $L$ is given by 
\begin{align}
L =3\rho_B\frac{\partial E_{\rm sym}(\rho_B)}{\partial \rho_B}\ .
\end{align}
Here, we assume that only the lightest nucleon exist in the nuclear matter at normal nuclear density. Then, this is reduced to
\begin{align}
L& =3\rho_B\frac{\partial E_{sym}}{\partial \rho_B}\bigg\vert_{\rho_B=\rho_0}
\\
& 
=3\rho_0 \Bigg[ \frac{1}{8} \left(\frac{2\pi^2}{k_FE_F}+\frac{g_\rho^2}{m_\rho^2}\right) - \frac{ \pi^2 \left( 2k_F^2+\left( m_N^\ast \right)^2 \right)}{ 24 k_F E_F^3 } \Bigg] \ ,
\end{align}
where 
$E_F=\bar{\mu}_B$
is the Fermi energy of the lightest nucleon, and 
$k_F$ is the Fermi momentum, 
$k_F = \sqrt{ \bar{\mu}_B^2 - \left( m_N^{\ast} \right)^2 }$.

\subsection{Neutron Star Matter and Tidal Deformability}

In this subsection, we construct neutron star matter based on the model introduced in 
section~\ref{sec: model}.
Here we assume that there are no hyperons and quarks in the matter constructed below.

For constructing the neutron star matter, we introduce the electron and muon into matter, and require conditions for the charge neutrality and the beta equilibrium.
The charge neutrality condition is written as
\begin{equation}
\sum_{i} \rho_p^{(i)} = \rho_e + \rho_\mu \ ,
\end{equation}
where $\rho_p^{(i)}$ is given in Eq.~(\ref{each density}).
The electron density and the muon density are given by
\begin{align}
\rho_l = 2 \int\frac{d^3k}{(2\pi)^3} \, \theta \left( \mu_l - E_l \right) \ , \quad (l = e, \mu) \ ,
\label{lepton density}
\end{align}
where $\mu_l$ is the corresponding chemical potential and 
\begin{equation}
E_l = \sqrt{ k^2 + m_l^2} \ ,
\end{equation}
with $m_l$ being the lepton mass.  Here we assume that the lepton masses in the neutron star matter are the same as those in vacuum.
The chemical potentials for leptons satisfy the chemical equilibrium conditions:
\begin{equation}
- \mu_I = \mu_e = \mu_\mu \ . 
\end{equation}

The mass and radius of neutron star are determined by solving Tolman-Oppenheimer-Volkov (TOV) equations~\cite{Tolman:1939jz,Oppenheimer:1939ne} given by
\begin{align}
\frac{dP(r)}{dr}= & -\frac{\epsilon(r)+P(r)}{r(r-2M(r))}\left[M(r)+4\pi^2r^3P(r)\right] \ ,
\notag \\
\frac{dM(r)}{dr}= & 4\pi^2 r^2 \epsilon(r) \ .
\label{TOV eq}
\end{align}
The solution of the above TOV equations determines the radius $R$ and the mass $M$ of the neutron star as
\begin{align}
P(r=R)=0 \ , \quad M = M(r = R) \ .
\end{align}

The dimenstionless tidal deformability is defined as~\cite{Hinderer:2007mb,Hinderer:2009ca,Malik:2018zcf}
\begin{align}
\Lambda=\frac{2}{3}k_2C^{-5} \ , 
\end{align}
where $C=M/R$ is the compactness parameter and $k_2$ is the tidal Love number calculated by
\begin{align}
k_2=& \frac{8C^5}{5}(1-2C)^2[2+2C(y_R-1)-y_R] \notag\\
& \ \ \times \Bigg[ 2C \left\{ 6-3y_R+3C(5y_R-8) \right\} \notag\\
& \quad {}+4C^3 \left\{ 13-11y_R+C(3y_R-2)+2C^2(1+y_R) \right\} \notag\\
& \quad {}+3(1-2C)^2\left\{ 2-y_R+2C(y_R-1){\rm ln}(1-2C)\right\}  \Bigg]^{-1} \ .
\end{align}
In this expression, the quantity $y_R=y(r=R)$ is obtained by solving the following differential equation:
\begin{align}
r\frac{dy(r)}{dr}+y(r)^2+y(r)F(r)+r^2+r^2Q(r)=0 \ ,
\end{align}
where
\begin{align}
F(r) = & \frac{r-4\pi r^3(\epsilon(r)-P(r))}{r-2M(r)} \ , \notag\\
Q(r)= & \frac{4\pi r(5\epsilon(r)+9P(r)+\frac{\epsilon(r)+P(r)}{\partial P(r)/\partial \epsilon(r)}-\frac{6}{4\pi r^2})}{r-2M(r)} \notag\\
& {} -4[\frac{M(r)+4\pi r^3P(r)}{r^2-2M(r)r}]^2 \ .
\end{align}

The gravitational wave GW170817 is measured from a binary neutron star merger, so that it gives an constraint to the binary dimensionless tidal deformability defined as
\begin{align}
\tilde{\Lambda}=\frac{16}{3}\frac{(M_1+12M_2)M_1^4\Lambda_1+(M_2+12M_1)M_2^4\Lambda_2}{(M_1+M_2)^5} \ .
\end{align}
where $\Lambda_{i}$ ($i=1,2$) is the tidal deformability of each neutron star.

\section{Numerical analysis}
\label{sec: numerical}

\begin{figure}[H]
\begin{center}
\includegraphics[clip,width=9.0cm]{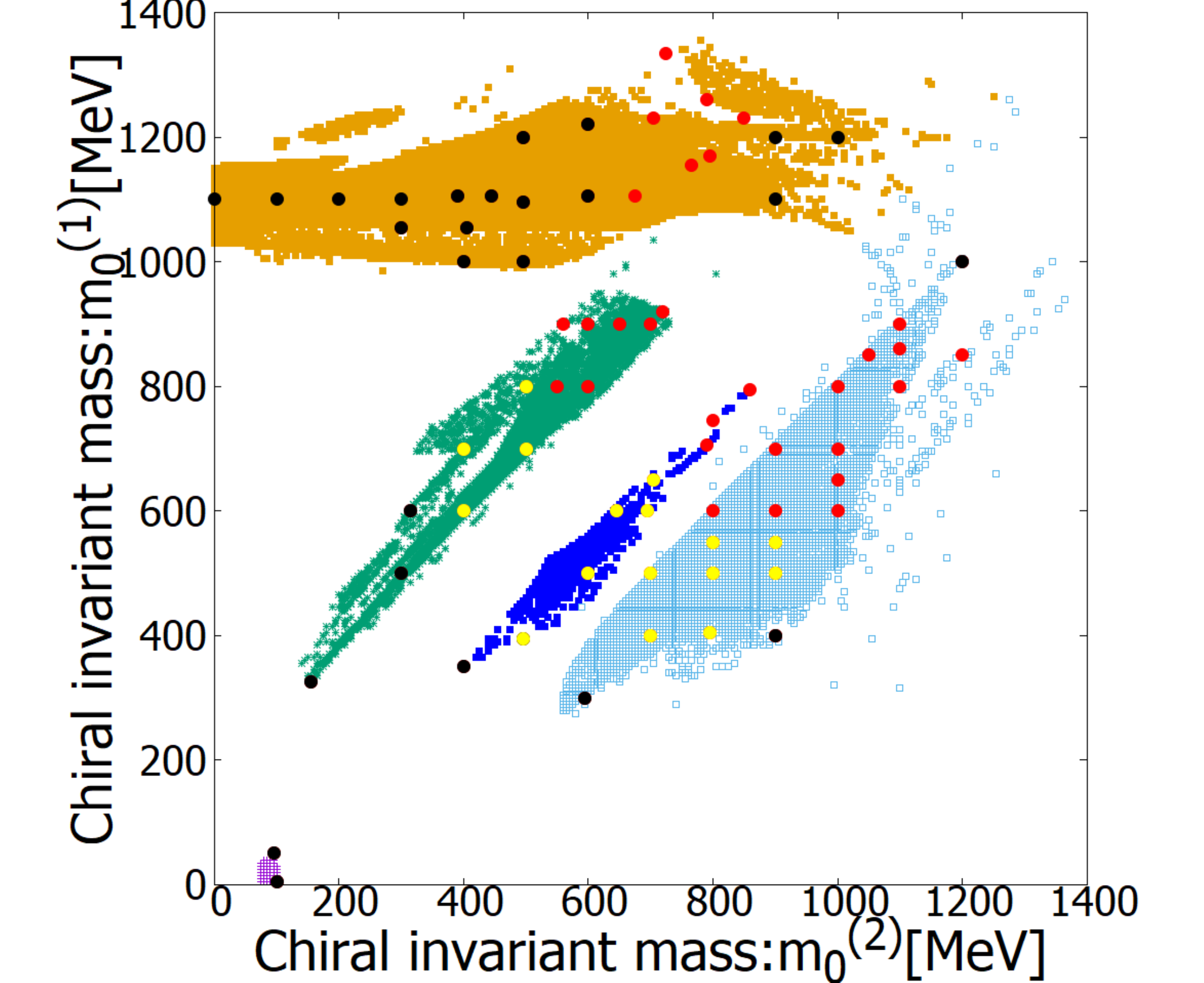}	
\end{center}
\caption{Allowed region for two chiral invariant masses. Painted regions indicate the solutions which reproduce the physical inputs at vacuum shown in Table~\ref{tab:vacuum inputs} as obtained in Ref.~\cite{Yamazaki:2018stk}.
For combinations of two chiral invariant masses indicated by black points, the saturation properties of normal nuclear matter are not satisfied (see subsection~\ref{subsec: SP}).
For the yellow points the obtained tidal deformability does not satisfy the constraint obtained by GW170817 (see subsection~\ref{subsec: TD}).
}
\label{fig: invariant masses}
\end{figure}

In Ref.~\cite{Yamazaki:2018stk}, we determined ten couplings of nucleons to scalar and pseudoscalar mesons from ten physical inputs shown in Table~\ref{tab:vacuum inputs}
for fixed values of two chiral invariant masses and the pion decay constant $f_\pi = 92.4$\,MeV.
\begin{table}[H]
\begin{center}
\begin{tabular}{|l|c|c|c|} \hline
& Mass & Width[$\Gamma_{N^*\to N\pi}$] & axial charge
\\ \hline \hline
    $N(939)$ & $939$ & - 			& $1.272$ \\
    $N(1440)$ & $1430$ & $228$ & -\\
    $N(1535)$ & $1535$ & $68$ & $-0.25$-$0.25$ [lat]\\ 
    \multirow{2}{*}{$N(1650)$} & \multirow{2}{*}{$1655$} & $84$ [to $N(939)]$ & \multirow{2}{*}{$0.55$ [lat]}\\	
	&		&	$22$ [to $N(1440)$]	&	\\	\hline
\end{tabular}
\end{center}
\caption[]{Physical inputs used to determine the couplings of nucleons to scalar and pseudoscalar mesons. Units of masses and widths are in MeV. Note that we adopt the restriction of $-0.25 \le g_A(1535) \le 0.25$ from the lattice analysis~\cite{Takahashi:2008fy} showing $g_A(1535) \sim {\mathcal O}(0.1)$.
}\label{tab:vacuum inputs}
\end{table}
We showed that only certain combinations of two masses can reproduce physical inputs, and that the solutions are categorized into five groups as shown in Fig.~\ref{fig: invariant masses}.
In the following we shall show that some regions are excluded by requiring the saturation properties of normal nuclear matter and the tidal deformability constraint from the observation of GW170817~\cite{TheLIGOScientific:2017qsa,GBM:2017lvd,Abbott:2018exr}.
Here we assume that all the model parameters do not have any density dependence.

\subsection{Saturation Properties}
\label{subsec: SP}

In addition to the parameters determined in Ref.~\cite{Yamazaki:2018stk} at vacuum as explained at the beginning of this section, we use the masses of $\rho$ and $\omega$ mesons as
\begin{equation}
m_{\omega}=783\,\mbox{MeV}\ , \quad m_\rho = 775\,\mbox{MeV} \ .
\end{equation}
Then, for fixed values of two chiral invariant masses, we determine four parameters, $g_{\omega}$, $g_{\rho}$, $\lambda_4$ and $\lambda_6$ to reproduce the saturation density, the binding energy, the incompressibility and the symmetry energy at the normal nuclear density shown in Table~\ref{expvalue}.
\begin{table}[H]
\begin{center}
\begin{tabular}{|c|c|c|c|} 
\hline
$\rho_0$	&	$E_{{\rm bind},\rho_0}$ &	$K$		&	$E_{\rm sym}$ \\
\hline
0.16					&	-16					&	240			&	31
\\
\hline
\end{tabular}
\end{center}
\caption{Inputs values of the saturation saturation density $\rho_0$, the binding energy $E_{{\rm bind},\rho_0}$, the incompressibility $K$ and the symmetry energy $E_{\rm sym}$ at the normal nuclear density.
Unit of $\rho_0$ is in $\mbox{fm}^{-3}$ and those for others are in MeV.}
\label{expvalue}
\end{table}

We use the combinations of two chiral invariant masses as shown by points in Fig.~\ref{fig: invariant masses}, and found that the saturation properties cannot be reproduced for the combinations indicated by the black dots.
Let us exiplain the reason why we excluded the combinations where both the chiral invariant masses are small.
For this purpose, we plot the density dependences of the pressure and energy density for 
$(m_0^{(1)},m_0^{(2)})=(325,155)$\,MeV (black dashed curves) which we excluded, together with $(900,700)$\,MeV (red curves) which satisfies the saturation properties.
\begin{figure}[H]
\begin{center}
\includegraphics[clip,width=8.0cm]{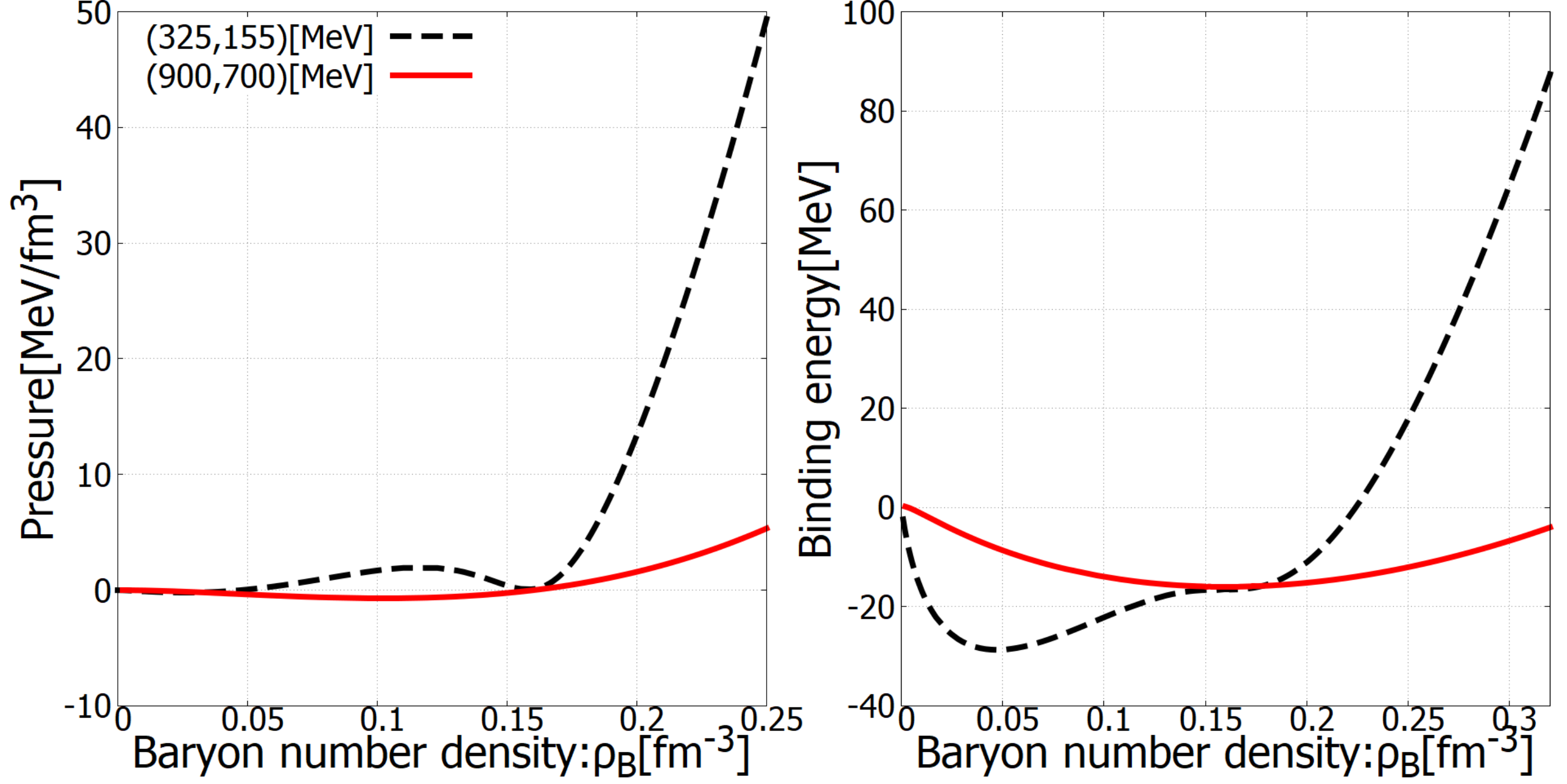}	
\end{center}
\caption{
Density dependences of the pressure (left panel) and the binding energy (right panel).
Black dashed curves are for $(m_0^{(1)},m_0^{(2)})=(325,155)$\,MeV, and the red solid curves are for $(900,700)$\,MeV.
}
\label{PEG2}
\end{figure}
Although both combinations satisfy $P = 0$ and $E/A - m_N = -16$\,MeV at $\rho_0 = 0.16\,\mbox{fm}^{-3}$, the black dashed curves show that the binding energy is not minimized at $\rho_0$ and there is a global minimum around $\rho_B \sim 0.05 \,\mbox{fm}^{-3}$. This implies that the matter at $\rho_0$ is not stable. 
This is because, for small chiral invariant masses, the coupling of the nucleon to $\sigma$ is large, and the attractive force is strong.

Similar situation occurs when one of two chiral invariant masses is small.
When both the chiral invariant masses are very large, on the other hand, 
the attractive force is too weak to keep the matter.

\subsection{Tidal Deformability}
\label{subsec: TD}

In this subsection, we construct the neutron star matter and calculate the tidal deformability using the formulas shown in the previous section.
Here we assume that there are no heperons and quarks in the matter.

\begin{figure}[H]
\begin{center}
\includegraphics[clip,width=9.0cm]{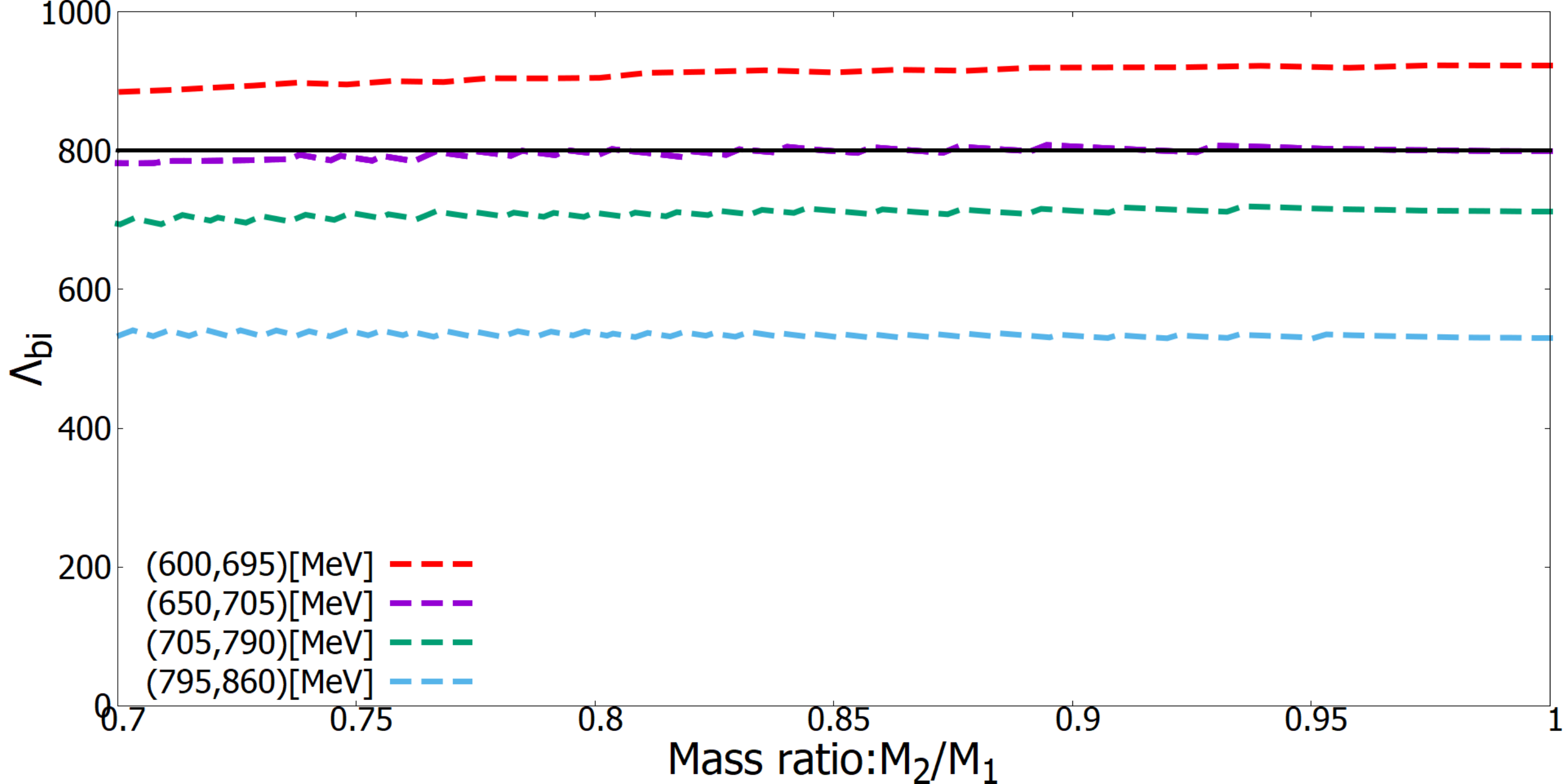}	
\end{center}
\caption{Binary dimensionless tidal deformability ($\Lambda_{\rm bi}\equiv \tilde{\Lambda}$ for several choices of two chiral invariant masses in Group 2. The horizontal axis shows the ratio of two masses of neutron stars.  Here we use the constraint of the mass ratio, $0.7 < M_2/M_1<1$~~\cite{TheLIGOScientific:2017qsa,Abbott:2018wiz}.
The dashed curves, from up to down, are for $(m_0^{(1)},m_0^{(2)}) = (600,695)$, $(650,705)$, $(705,790)$, $(795,860)$\,MeV, respectively.
The black solid line at $\tilde{\Lambda}=800$ is the upper bound of the constraint from the observation of 
GW170817~\cite{TheLIGOScientific:2017qsa,Abbott:2018wiz}.
 }
\label{BDTDG2}
\end{figure}

When we solve the TOV equation (\ref{TOV eq}), we use the Equation of State (EOS) obtained from the present model for $\rho_B > 0.1\,\mbox{fm}^{-3}$, while we use the EOS given in Refs.~\cite{Baym:1971pw,Sharma:2015bna} for $\rho_B < 0.1\,\mbox{fm}^{-3}$.
Then, we calculate the binary tidal deformability for the Chirp mass of
\begin{equation}
M_{\rm Chirp} = \frac{ \left( M_1 M_2 \right)^{3/5} }{ \left( M_1 + M_2 \right)^{1/5} } = 1.118 M_{\odot} \ , 
\end{equation}
where $M_{\odot}$ is the solar mass.
We show the resultant values of the binary tidal deformability  for several combinations of two chiral invariant masses for Group 2 in Fig.~\ref{BDTDG2}.
Here the dashed curves, from up to down, are for $(m_0^{(1)},m_0^{(2)}) = (600,695)$, $(650,705)$, $(705,790)$ and $(795,860)$\,MeV, respectively.
We also plot the black sold line at $\tilde{\Lambda}=800$ which we regard as 
the upper bound of the constraint from the observation of GW170817~\cite{TheLIGOScientific:2017qsa,Abbott:2018wiz}.
Figure~\ref{BDTDG2} shows that the $\tilde{\Lambda}$ become smaller for larger chiral invariant mass. 
This is because the EOS becomes softer for larger chiral invariant mass.

As a result, the constraint $\tilde{\Lambda} < 800$ excludes the region where chiral invariant masses are small.
For example, the combination $(m_0^{(1)},m_0^{(2)}) = ( 600 ,695 )$\,MeV is excluded as one can see easily in Fig.~\ref{BDTDG2}.

\begin{figure}[H]
\begin{center}
\includegraphics[clip,width=9.0cm]{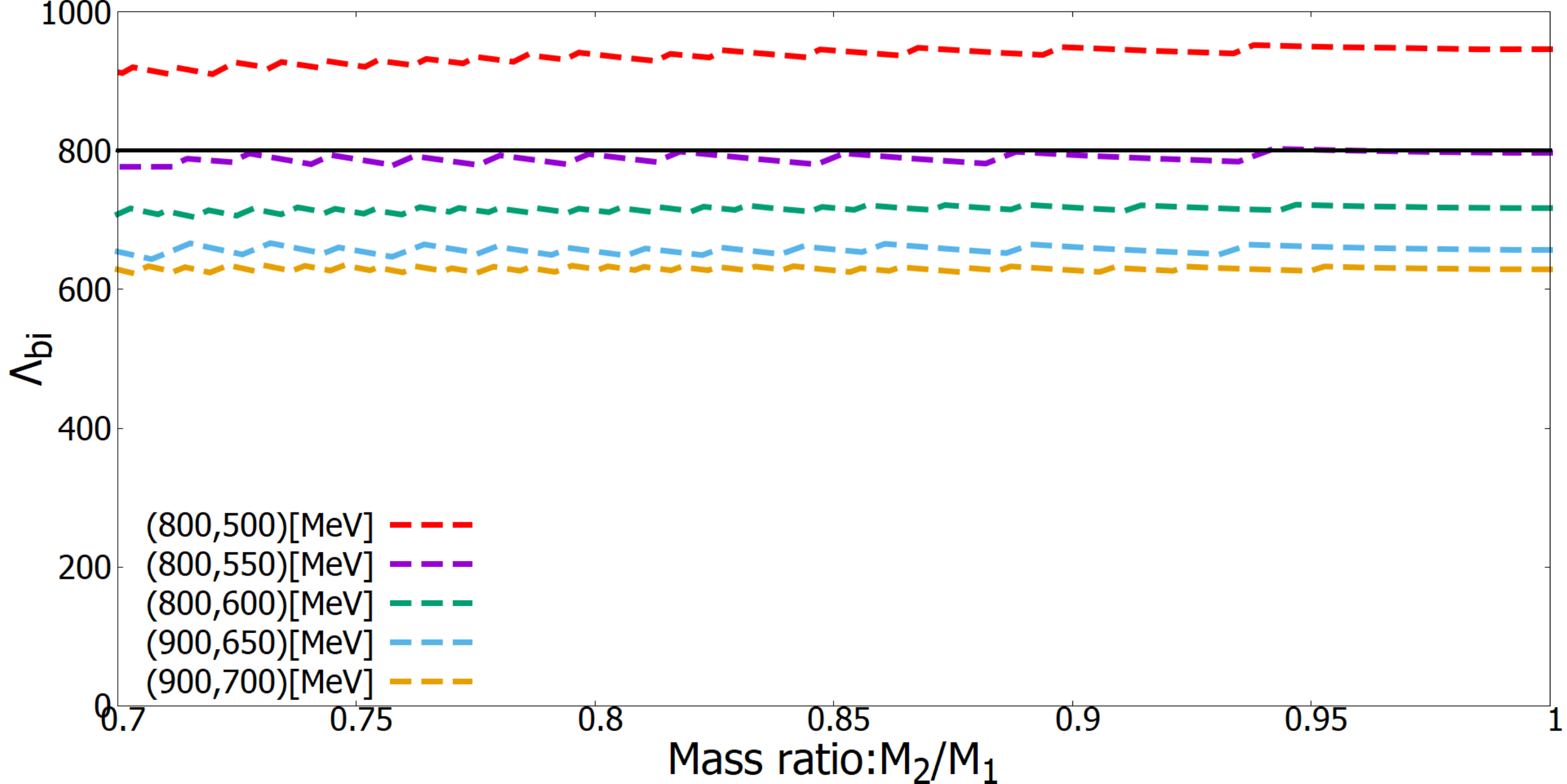}	
\end{center}
\caption{Binary dimensionless tidal deformability ($\Lambda_{\rm bi}\equiv \tilde{\Lambda}$ for several choices of two chiral invariant masses in Group 3.
The dashed curves, from up to down, are for $(m_0^{(1)},m_0^{(2)}) = (800,500)$, $(800,550)$, $(800,600)$, $(900,650)$, $(900,700)$\,MeV, respectively.
}
\label{BDTDG3}
\end{figure}

\begin{figure}[H]
\begin{center}
\includegraphics[clip,width=9.0cm]{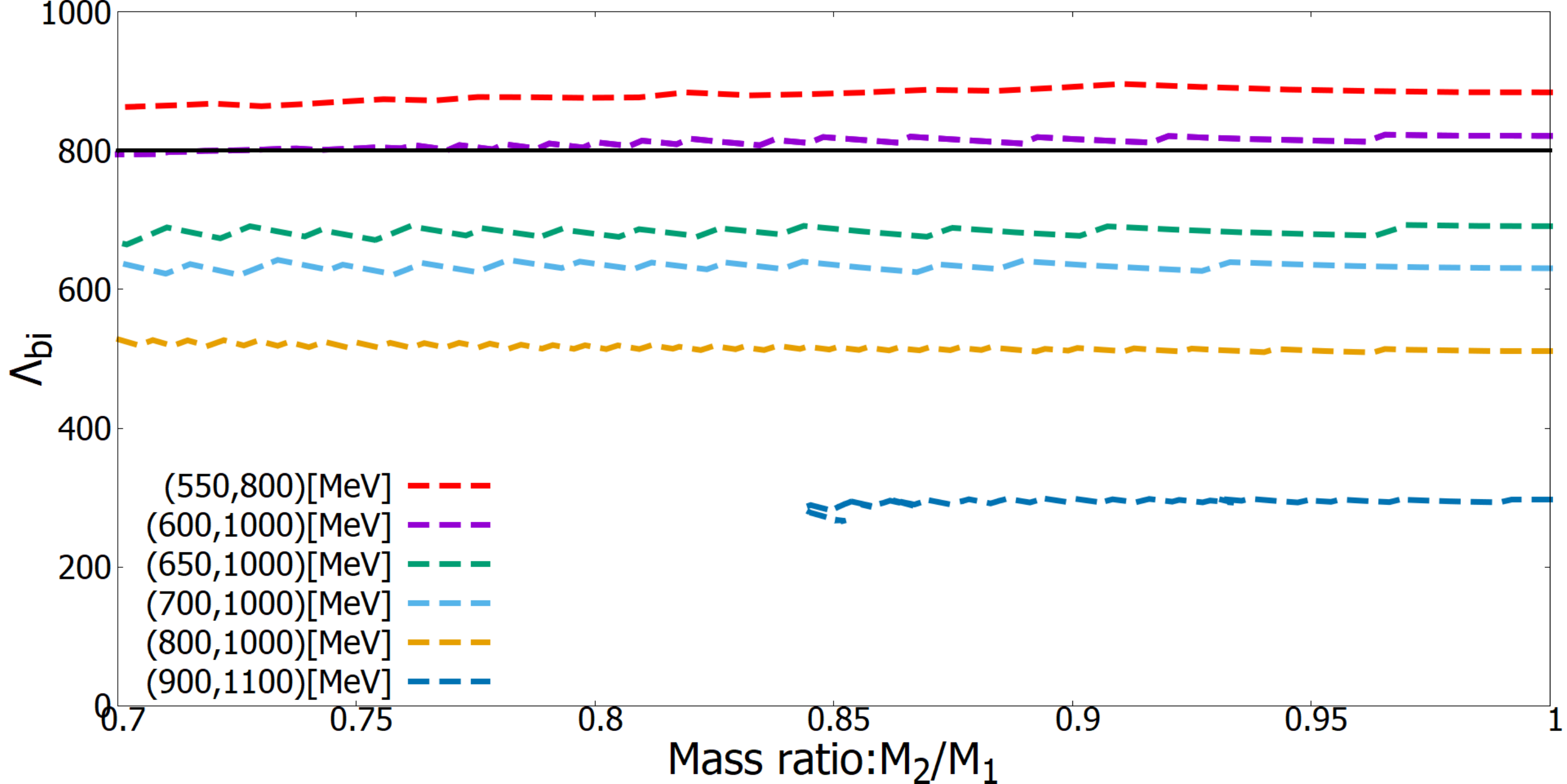}	
\end{center}
\caption{Binary dimensionless tidal deformability ($\Lambda_{\rm bi}\equiv \tilde{\Lambda}$ for several choices of two chiral invariant masses in Group 4.
The dashed curves, from up to down, are for $(m_0^{(1)},m_0^{(2)}) = (550,800)$, $(600,1000)$, $(650,1000)$, $(700,1000)$, $(800,1000)$, $(900,1000)$\,MeV, respectively.
}
\label{BDTDG4}
\end{figure}

\begin{figure}[H]
\begin{center}
\includegraphics[clip,width=9.0cm]{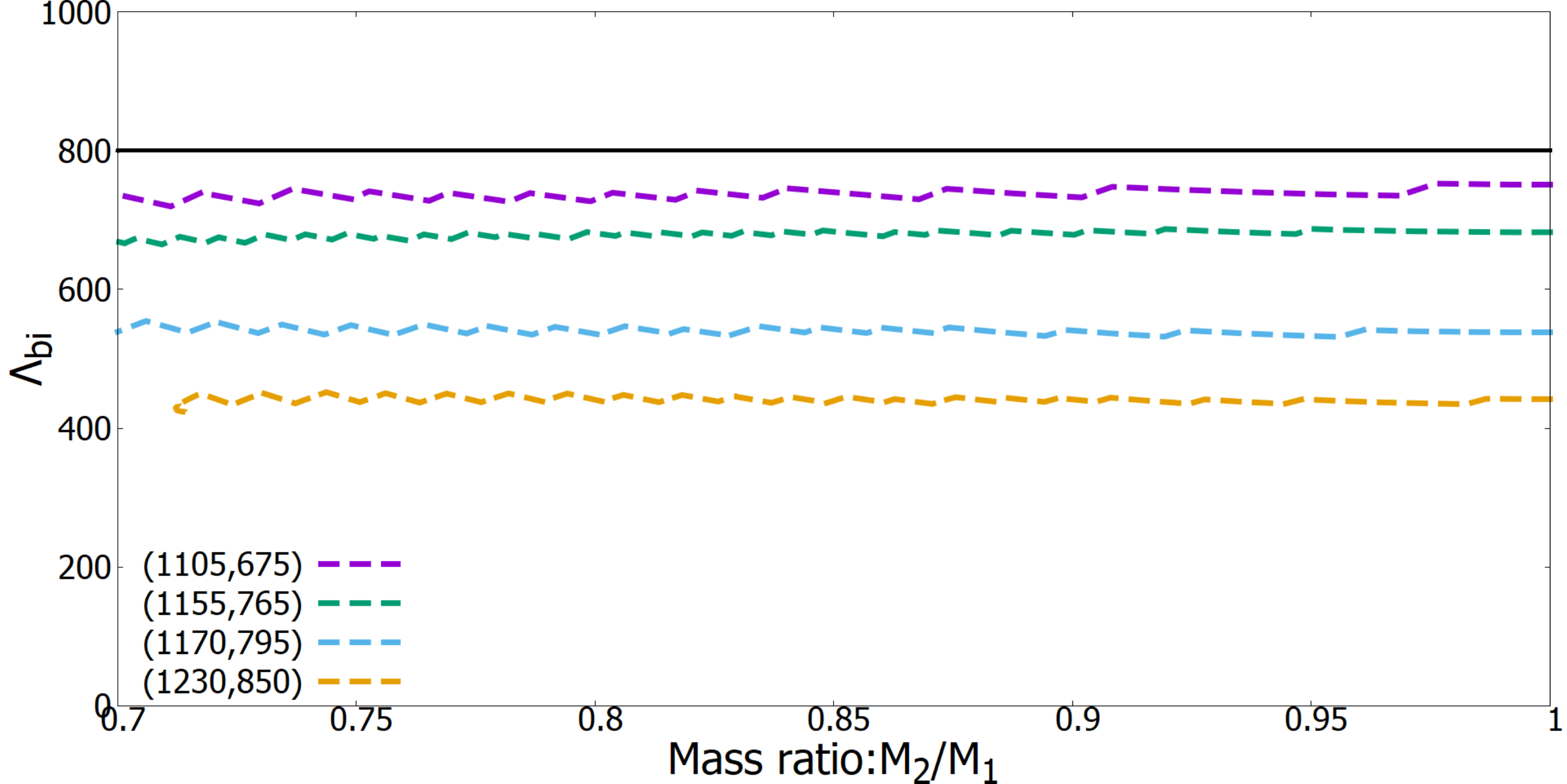}	
\end{center}
\caption{Binary dimensionless tidal deformability ($\Lambda_{\rm bi}\equiv \tilde{\Lambda}$ for several choices of two chiral invariant masses in Group 5.
The dashed curves, from up to down, are for $(m_0^{(1)},m_0^{(2)}) = (1105,675)$, $(1155,765)$, $(1170,795)$, $(1230,850)$\,MeV, respectively.
}
\label{BDTDG5}
\end{figure}

We next show the predicted $\tilde{\Lambda}$ for Groups 3, 4, 5 in Figs.~\ref{BDTDG3}, \ref{BDTDG4}, \ref{BDTDG5}, respectively.
One can easily see that $\tilde{\Lambda}$ is larger for smaller chiral invariant masses, and 
the combinations $(m_0^{(1)},m_0^{(2)}) = ( 800 ,500 )$, $(550,800)$\,MeV are excluded.
We summarize the results in Fig.~\ref{fig: invariant masses}, where red points show that $\tilde{\Lambda} < 800$ is satisfied while yellow points show that the combination of the chiral invariant masses is excluded.
From this, we conclude that the chiral invariant masses must be larger than about $600$\,MeV to satisfy the tidal deformability constraint from the observation of GW170817.

\subsection{$M$-$R$ Relation}

In this subsection, we show our results for the $M$-$R$ relation and the central density.

\begin{figure}[H]
\begin{center}
\includegraphics[clip,width=9.0cm]{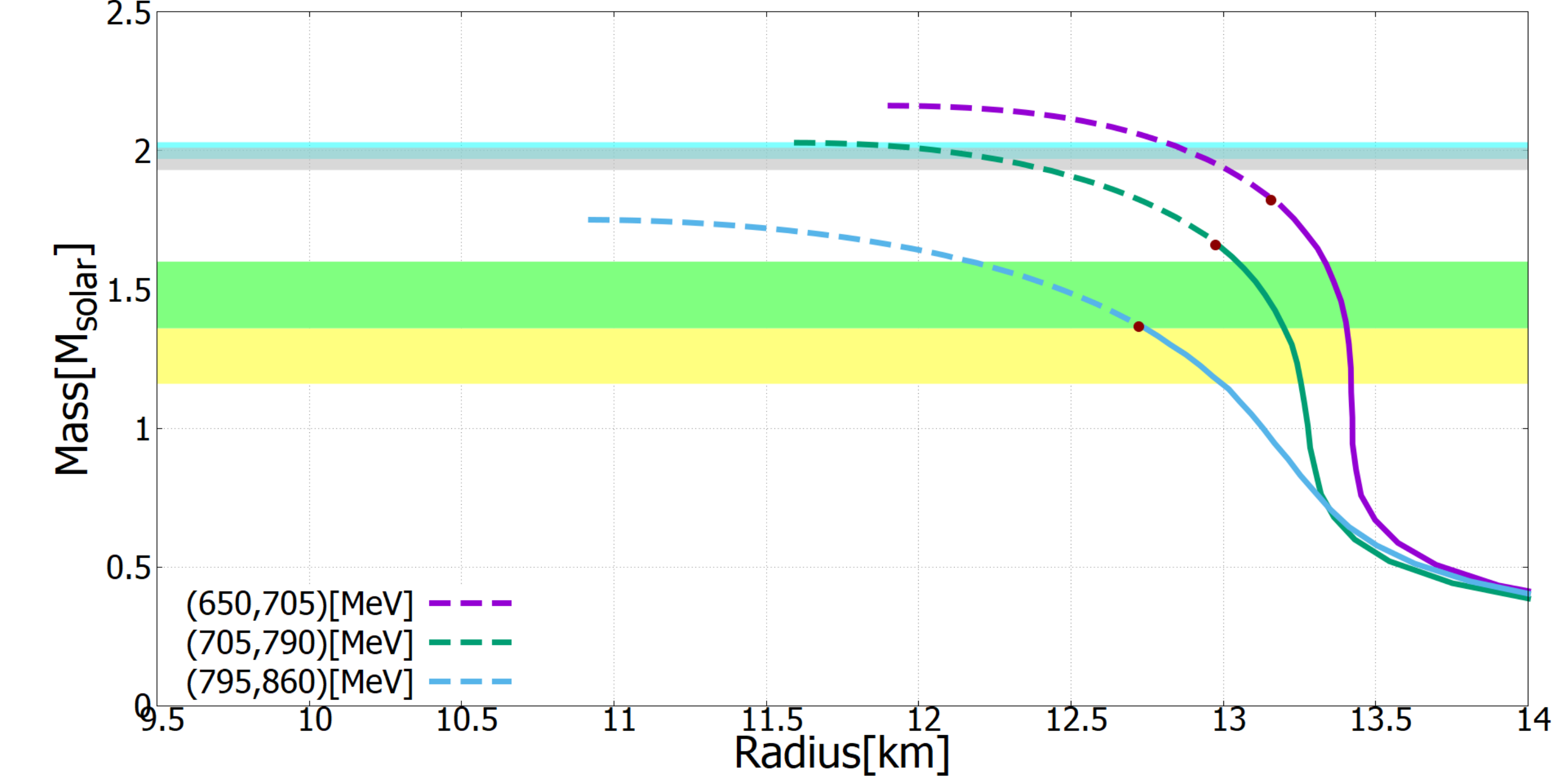}	
\end{center}
\caption{Relation between the mass and radius of neutron stars for several combinations of two chiral invariant masses in Group 2.
The curves are for $(m_0^{(1)},m_0^{(2)}) = (650,705)$, $(705,790)$, $(795,860)$\,MeV from up to down.
Solid curves imply that the central density $\rho_c$ is smaller than the three times of normal nuclear matter density, $\rho_c < 3 \rho_0$, and the dashed curves are for $\rho_c > 3 \rho_0$.
Dots on the curves express that the central density is three times of the normal nuclear density, $\rho_c=3\rho_0$.
}
\label{MRG2}
\end{figure}
\begin{figure}[H]
\begin{center}
\includegraphics[clip,width=9.0cm]{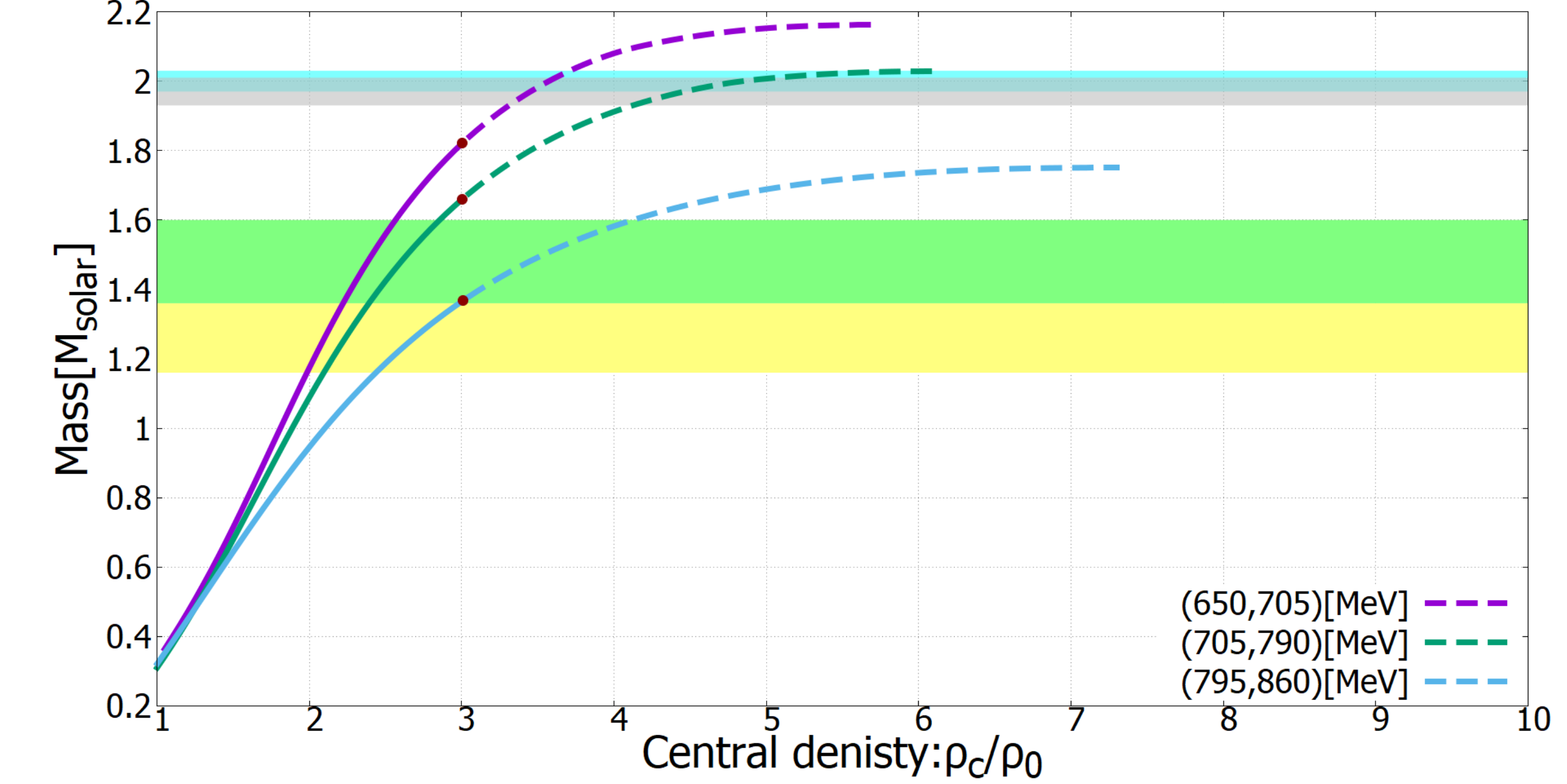}	
\end{center}
\caption{Relation between the mass and central density of neutron stars for several combinations of two chiral invariant masses in Group 2.
The curves are for $(m_0^{(1)},m_0^{(2)}) = (650,705)$, $(705,790)$, $(795,860)$\,MeV from up to down.
}
\label{MDcG2}
\end{figure}
We plot the relation between neutron star mass and radius 
in Fig.~\ref{MRG2} and the relation between the mass and the central density in Fig.~\ref{MDcG2} for several combinations of two chiral invariant masses $(m_0^{(1)},m_0^{(2)})$ in Group 2 which satisfy the constraint of the tidal deformability.
Here, solid curves imply that the central density $\rho_c$ is smaller than the three times of normal nuclear matter density, $\rho_c < 3 \rho_0$, and the dashed curves are for $\rho_c > 3 \rho_0$.
One may say that 
the combination 
$(m_0^{(1)},m_0^{(2)}) = (795,860)$\,MeV is excluded, since 
the present prediction does not reproduce the super-heavy neutron stars with 
$2M_{\odot}$ mass~\cite{Demorest:2010bx,Antoniadis:2013pzd}.
However, we note that, in this study, we assume that the core of neutron star is constructed by protons, neutrons, and leptons only and that none of hyperons, meson condensation and quark degrees appear.
We expect that the predictions shown by solid curves are not changed significantly, but
those by dashed curves will be changed.
Then, although the present prediction for $(m_0^{(1)},m_0^{(2)}) = (795,860)$\,MeV does not reproduce the super-heavy neutron stars with 
$2M_{\odot}$ mass~\cite{Demorest:2010bx,Antoniadis:2013pzd}, it will be changed by e.g., including effects of quark degrees~(See e.g., 
Refs.~\cite{Masuda:2012kf,Masuda:2012ed,Dexheimer:2012eu,%
Masuda:2015wva,Masuda:2015kha,%
Kojo:2014rca,Kojo:2015nzn,Baym:2017whm,Mukherjee:2017jzi,Wu:2018kww}.).

\begin{figure}[H]
\begin{center}
\includegraphics[clip,width=9.0cm]{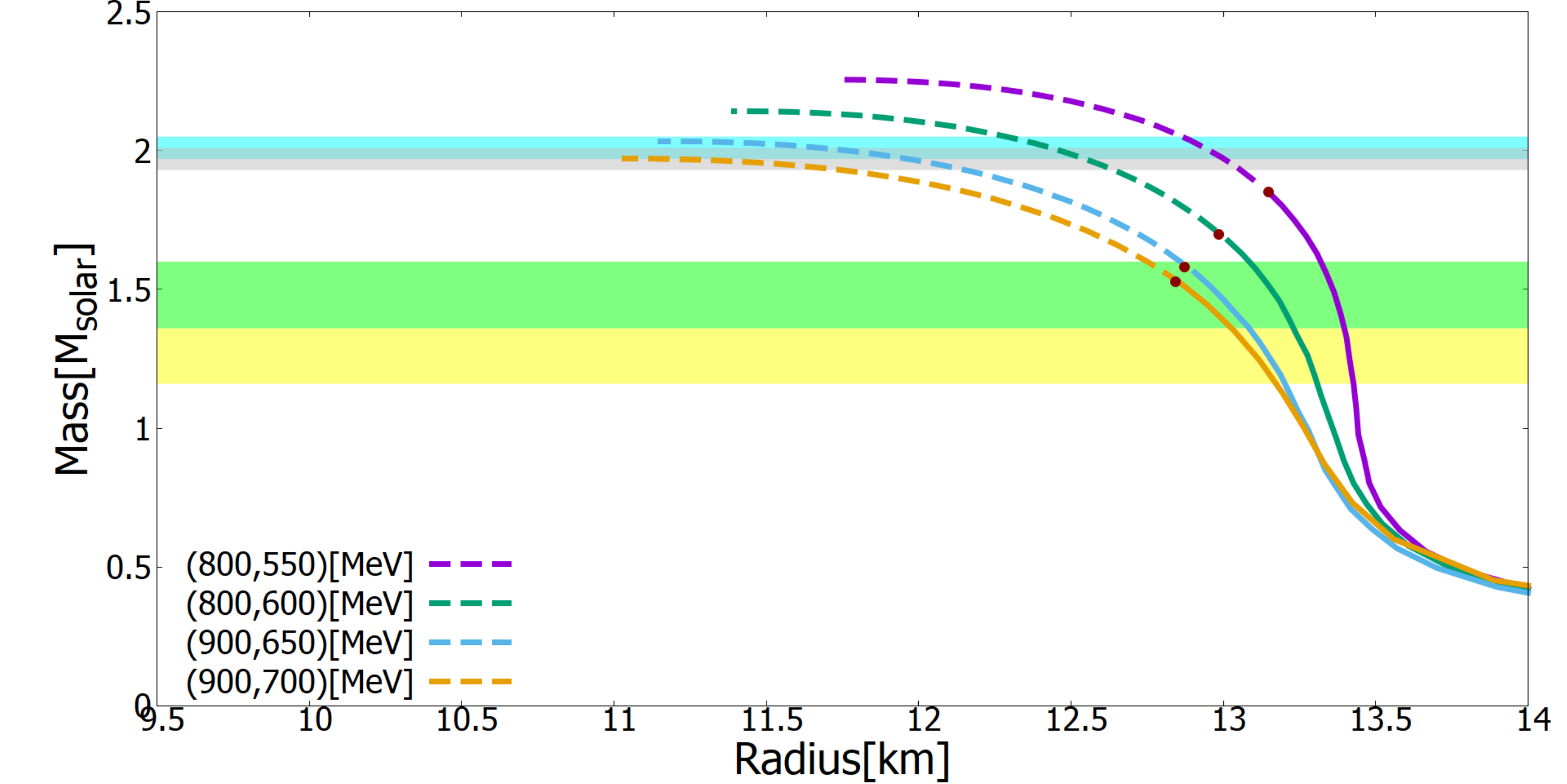}	
\end{center}
\caption{Relation between the mass and radius of neutron stars for several combinations of two chiral invariant masses in Group 3.
The curves are for $(m_0^{(1)},m_0^{(2)}) = (800,550)$, $(800,600)$, $(900,600)$, $(900,700)$\,MeV from up to down.
}
\label{MRG3}
\end{figure}

\begin{figure}[H]
\begin{center}
\includegraphics[clip,width=9.0cm]{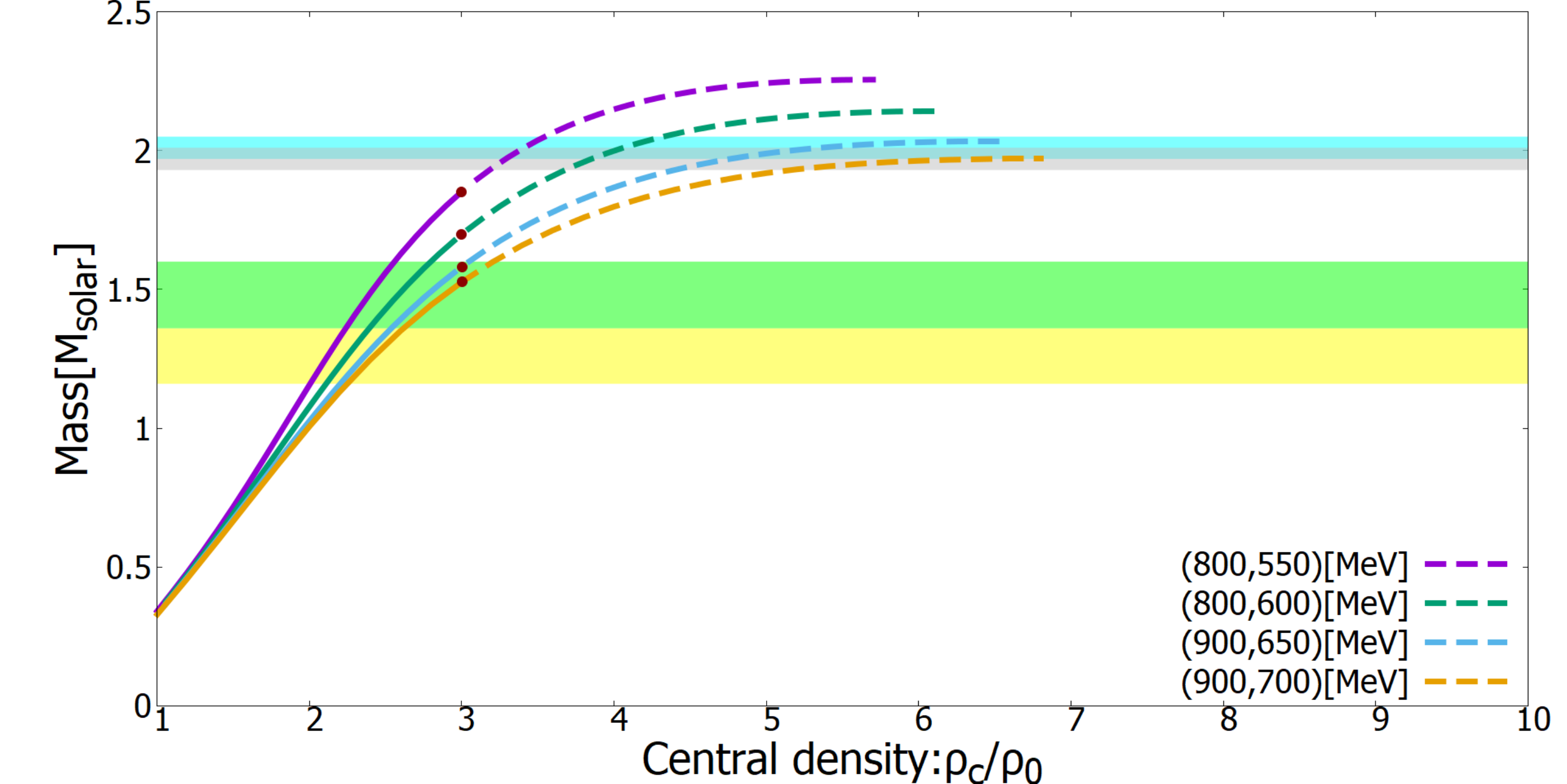}	
\end{center}
\caption{Relation between the mass and central density of neutron stars for several combinations of two chiral invariant masses in Group 3.
The curves are for $(m_0^{(1)},m_0^{(2)}) = (800,550)$, $(800,600)$, $(900,600)$, $(900,700)$\,MeV from up to down.
}
\label{MDcG3}
\end{figure}

\begin{figure}[H]
\begin{center}
\includegraphics[clip,width=9.0cm]{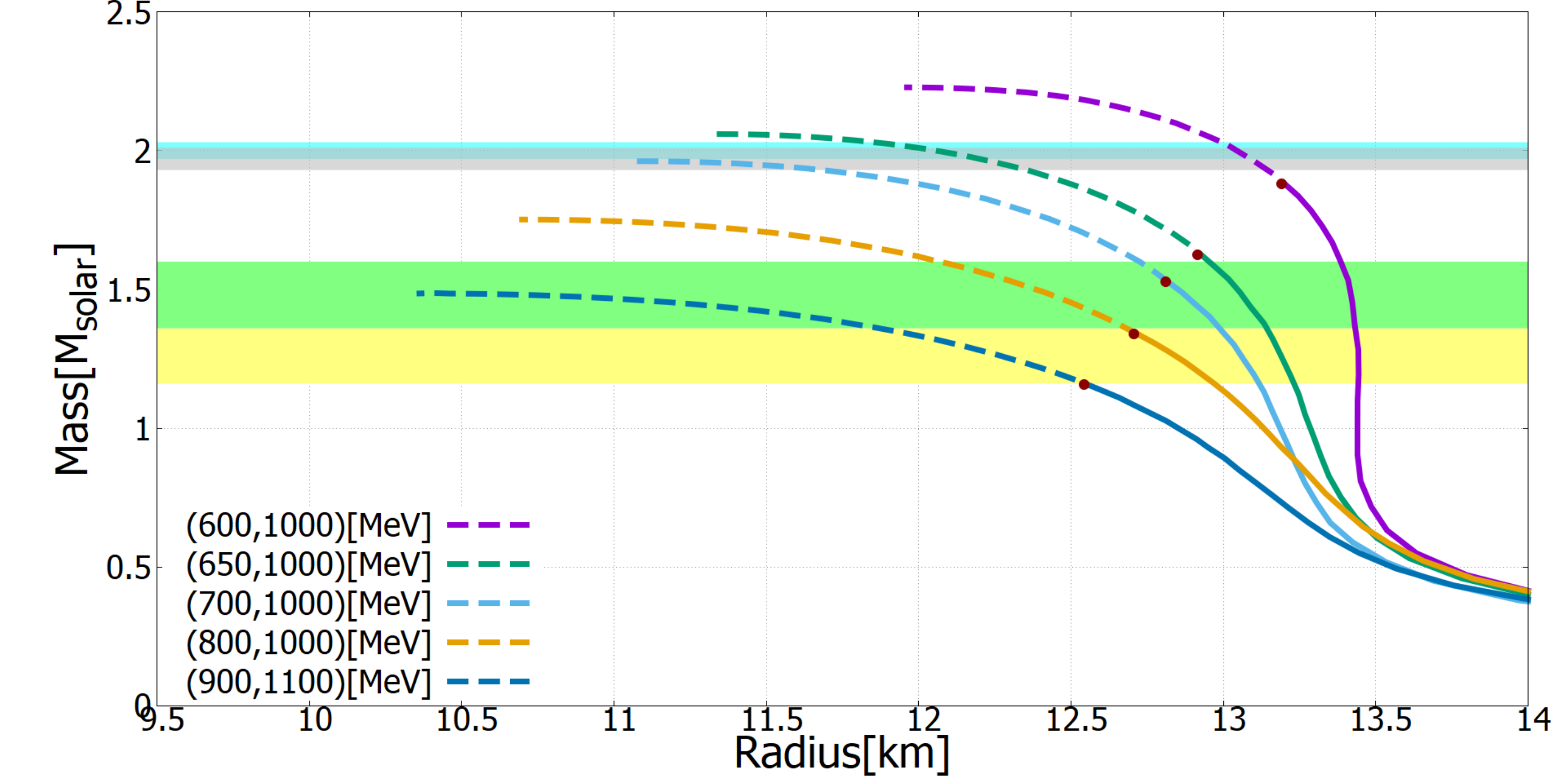}	
\end{center}
\caption{Relation between the mass and radius of neutron stars for several combinations of two chiral invariant masses in Group 4.
The curves are for $(m_0^{(1)},m_0^{(2)}) = (600,1000)$, $(650,1000)$, $(700,1000)$, $(800,1000)$, $(900,1100)$\,MeV from up to down.
}
\label{MRG4}
\end{figure}

\begin{figure}[H]
\begin{center}
\includegraphics[clip,width=9.0cm]{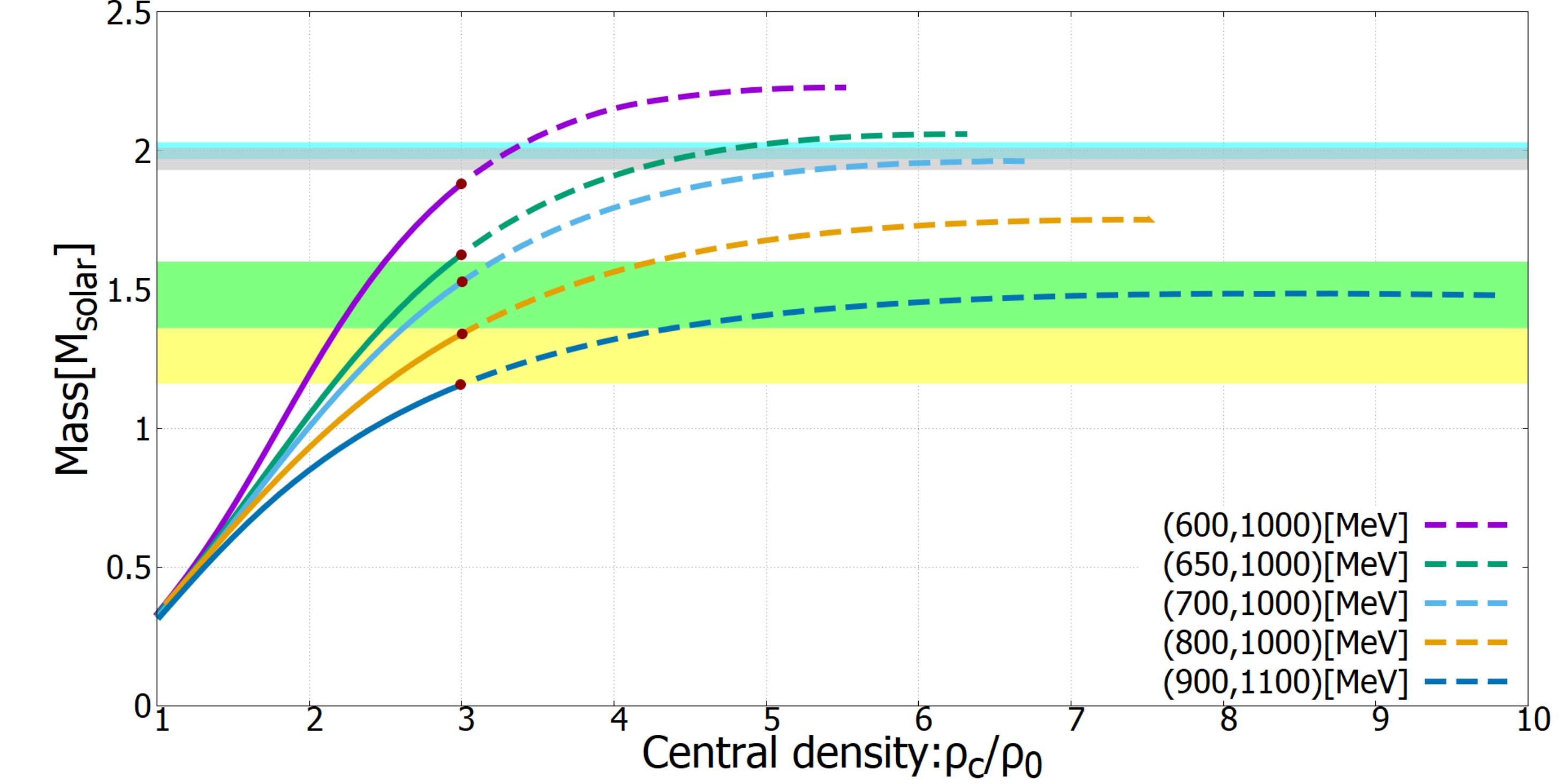}	
\end{center}
\caption{Relation between the mass and central density of neutron stars for several combinations of two chiral invariant masses in Group 4.
The curves are for $(m_0^{(1)},m_0^{(2)}) = (600,1000)$, $(650,1000)$, $(700,1000)$, $(800,1000)$, $(900,1100)$\,MeV from up to down.
}
\label{MDcG4}
\end{figure}

\begin{figure}[H]
\begin{center}
\includegraphics[clip,width=9.0cm]{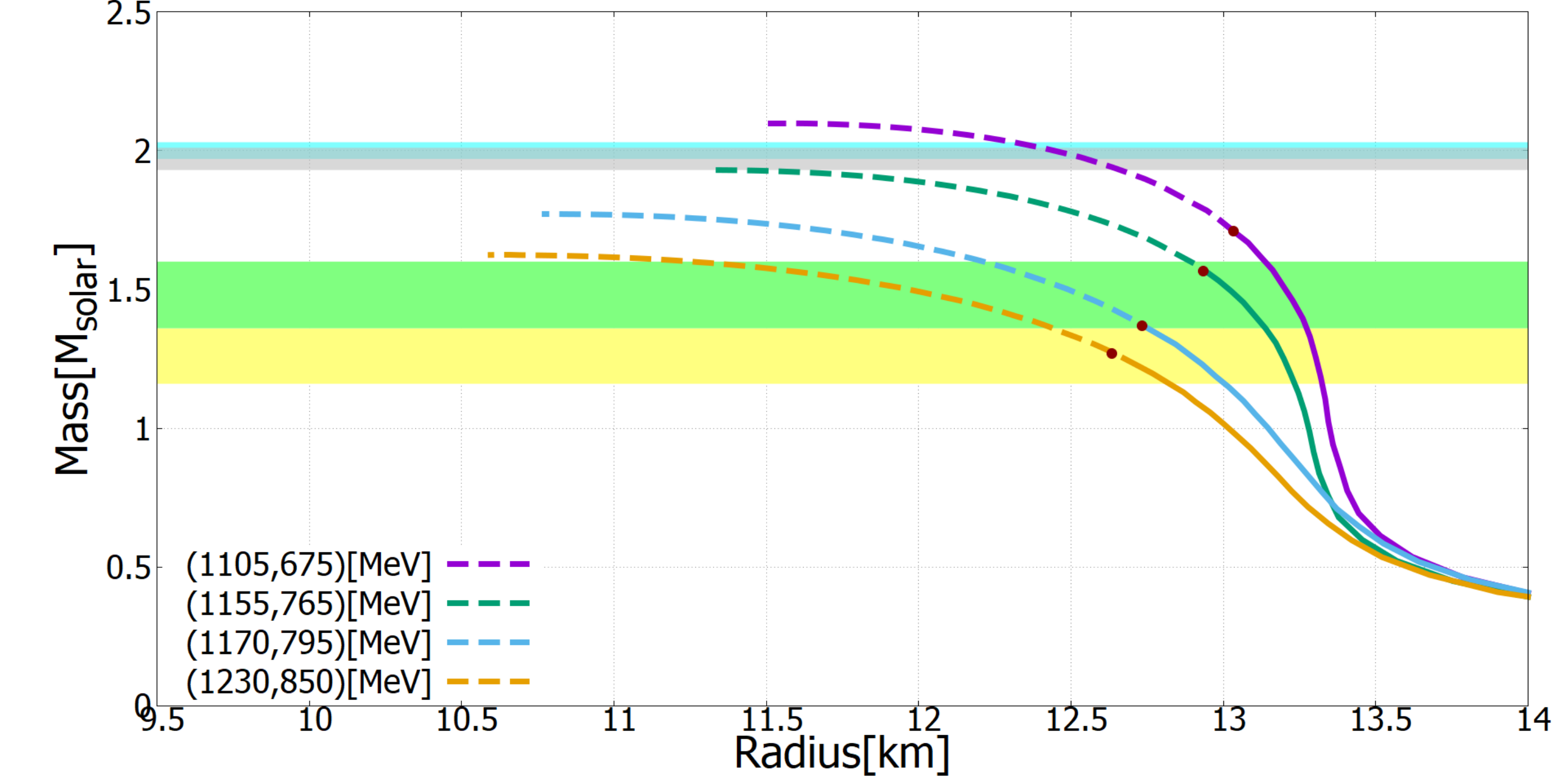}	
\end{center}
\caption{Relation between the mass and radius of neutron stars for several combinations of two chiral invariant masses in Group 5.
The curves are for $(m_0^{(1)},m_0^{(2)}) = (1105,675)$, $(1155,765)$, $(1170,795)$, $(1230,850)$\,MeV from up to down.
}
\label{MRG5}
\end{figure}

\begin{figure}[H]
\begin{center}
\includegraphics[clip,width=9.0cm]{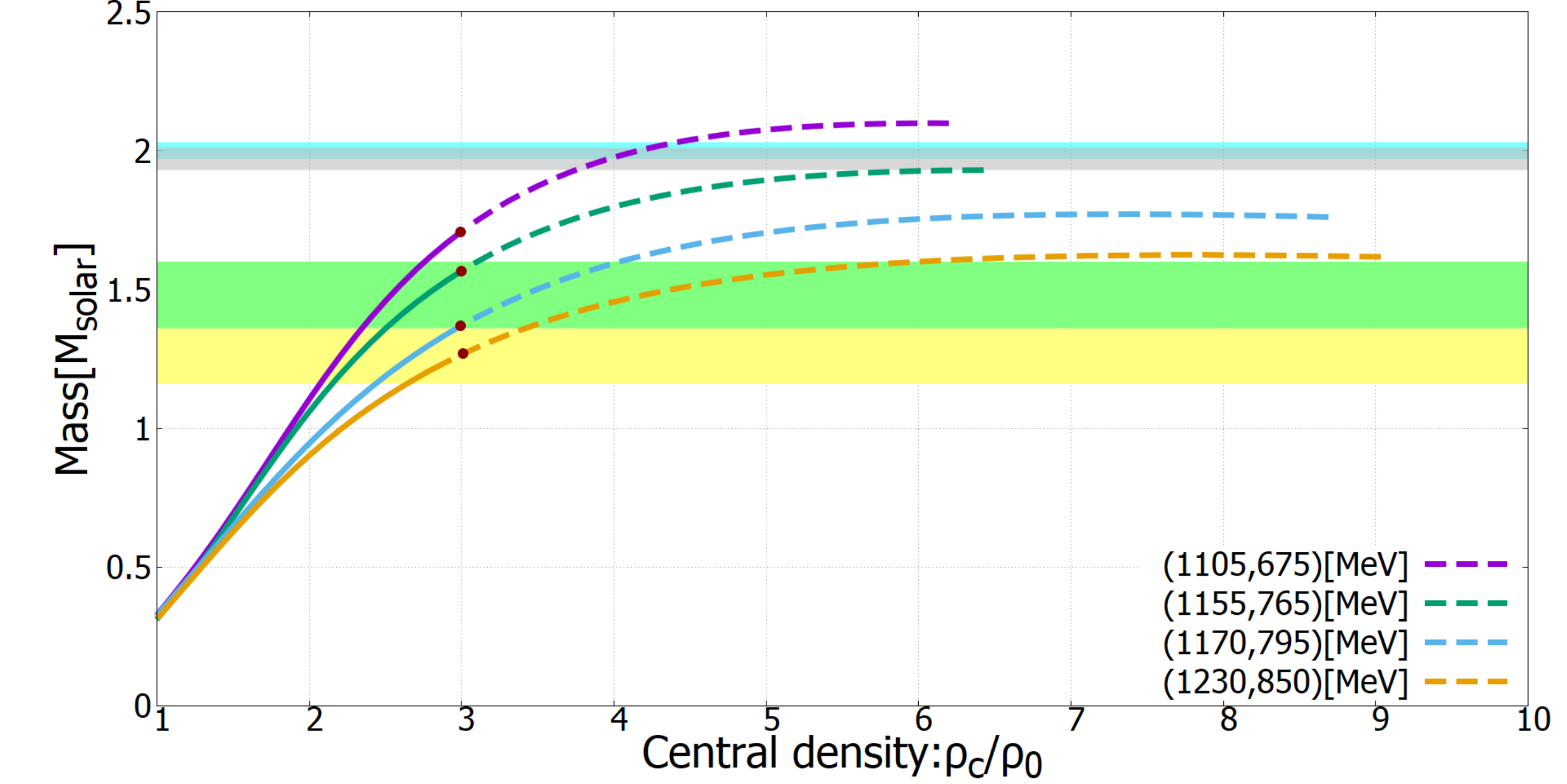}	
\end{center}
\caption{Relation between the mass and central density of neutron stars for several combinations of two chiral invariant masses in Group 5.
The curves are for $(m_0^{(1)},m_0^{(2)}) = (1105,675)$, $(1155,765)$, $(1170,795)$, $(1230,850)$\,MeV from up to down.
}
\label{MDcG5}
\end{figure}

We also list the relations for Groups 3 - 5 in Figs.~\ref{MRG3}-\ref{MDcG5}.
From these figures, we can see that the larger chiral invariant mass provides the softer EOS leading to smaller radius and lighter mass.
This is because the larger chiral invariant mass leads to the smaller repulsive interaction.

\subsection{Symmetry Energy and Slope Parameter}

\begin{figure}[H]
\begin{center}
\includegraphics[clip,width=9.0cm]{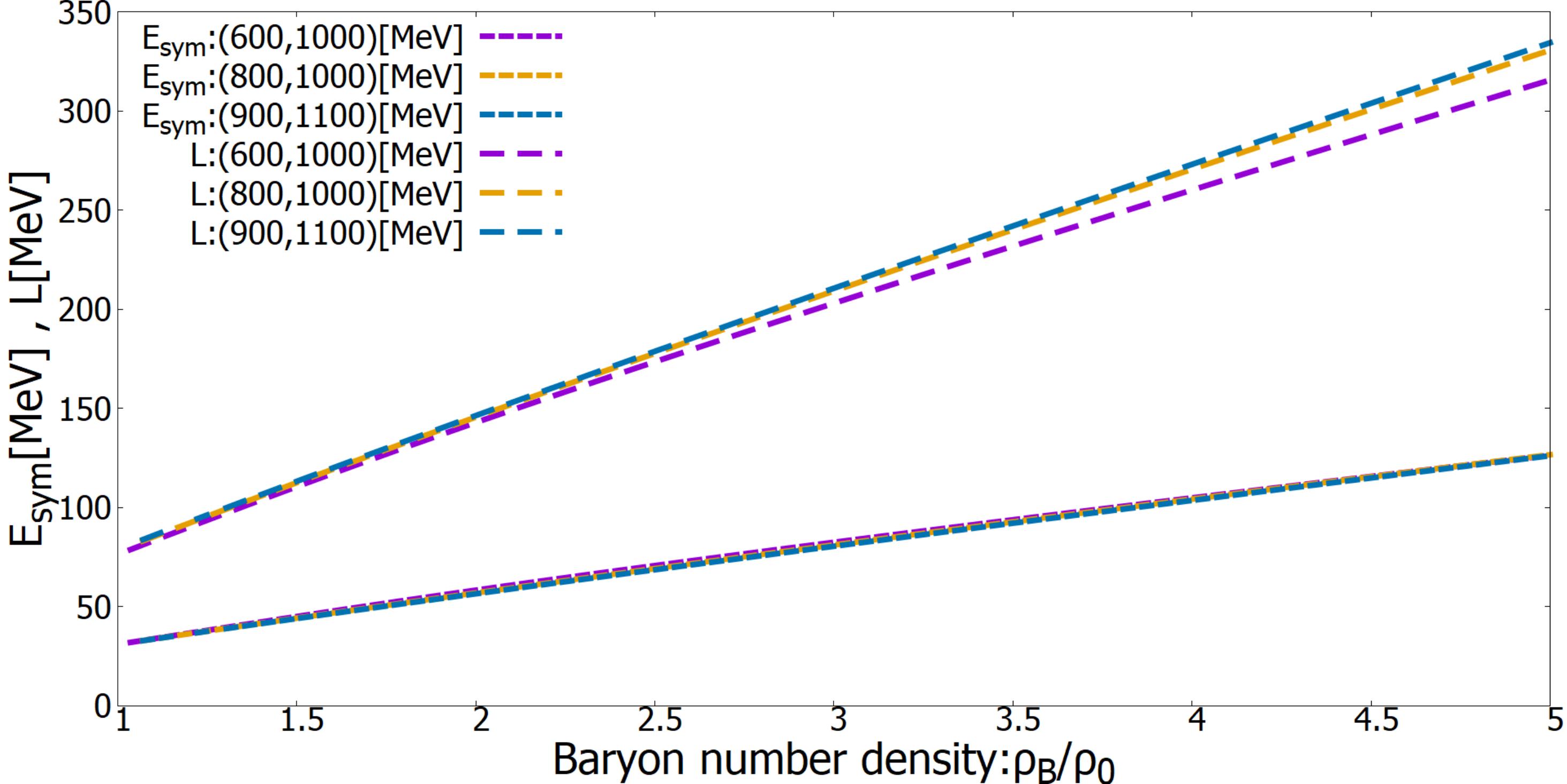}	
\end{center}
\caption{Predicted symmetry energy (lower three curves) and slope parameter (upper three curves) for $(m_0^{(1)},m_0^{(2)}) = (900,1100)$, $(800,1000)$, $(600,1000)$MeV, from up to down.
}
\label{fig: Esym}
\end{figure}

In this subsection, we calculate the symmetry energy and the slope parameter in high density matter. The plots for $(m_0^{(1)},m_0^{(2)}) = (900,1100)$, $(800,1000)$, $(600,1000)$MeV
are shown in Fig.~\ref{fig: Esym}.
We can see that three predictions are almost same, and we checked that other predictions are similar.
This is because, in Eq.~(\ref{eq:Esym}), the second term in the parenthesis is dominant and the symmetry energy is proportional to the baryon number density $\rho_B$. 
The slope parameters predicted here seems a little larger than the one constrained in Ref.~\cite{Kolomeitsev:2016sjl,Ohnishi:private}.
In the present model,
the large slope parameter causes the large radius of neutron stars.

\section{A Summary and Discussion}
\label{sec: summary}

In this paper, we constructed nuclear matter including neutron star matter from an extended parity doublet model which we developed in the previous publication~\cite{Yamazaki:2018stk}, where two sets of chiral representations having two chiral invariant masses are introduced.
We showed that some combinations of two chiral invariant masses are excluded by requiring the saturation properties of normal nuclear matter; the saturation density, the binding energy, the incompressibility, the symmetry energy.
Then, we found that more constraint to the chiral invariant masses is obtained from the tidal deformability determined by the observation of GW170817.
Out result shows that the chiral invariant masses are larger than about $600$\,MeV, which is consistent with the constraint obtained in Ref.~\cite{Marczenko:2018jui}.
We also showed predictions of the mass-radius relation of neutron stars, as well as the symmetry energy and the slope parameter at high density.

In the present analysis, some of our predictions do not reproduce the super-heavy neutron stars with $2M_{\odot}$~\cite{Demorest:2010bx,Antoniadis:2013pzd}.
However, we think that our model may not be applicable in the high-density region.
It might be changed by e.g., including effects of quark degrees~(See e.g., 
Refs.~\cite{Masuda:2012kf,Masuda:2012ed,Dexheimer:2012eu,%
Masuda:2015wva,Masuda:2015kha,%
Kojo:2014rca,Kojo:2015nzn,Baym:2017whm,Mukherjee:2017jzi,Wu:2018kww}.).

The slope parameters predicted here seems a little larger than the one constrained in Ref.~\cite{Kolomeitsev:2016sjl,Ohnishi:private}.
In the present model,
the large slope parameter causes the large
 radius of neutron stars, which are comparable with the results in e.g.,
Refs.~\cite{Most:2018hfd,Fattoyev:2017jql,Kim:2018aoi},
but larger than the ones in 
Refs.~\cite{Kojo:2014rca,Kojo:2015nzn,Baym:2017whm,Steiner:2010fz,Lattimer:2014sga,Togashi:2017mjp} .

In our model, we include the six-point interaction of the $\sigma$ meson. 
The existence of 
six point interaction might be driven by the violation of scale symmetry~\cite{Ma:2016nki}.

It is interesting to apply the present analysis to 
study the modification of spectrum of heavy-light mesons in dense matter
as done in e.g. Refs.~\cite{Harada:2016uca,Suenaga:2018kta}.

\acknowledgements

We would like to thank Youngman Kim, Young-Min Kim, Toru Kojo, Chang-Hwan Lee, Akira Ohnishi, Chihiro Sasaki for valuable discussions and comments.
The work of M.H. is supported in part by 
JPSP KAKENHI
Grant Number 16K05345.



\begin{thebibliography}{}


\bibitem{Detar:1988kn} 
  C.~E.~Detar and T.~Kunihiro,
  ``Linear $\sigma$ Model With Parity Doubling,''
  Phys.\ Rev.\ D {\bf 39}, 2805 (1989).
  doi:10.1103/PhysRevD.39.2805

\bibitem{Jido:1998av} 
  D.~Jido, Y.~Nemoto, M.~Oka and A.~Hosaka,
  ``Chiral symmetry for positive and negative parity nucleons,''
  Nucl.\ Phys.\ A {\bf 671}, 471 (2000)
  doi:10.1016/S0375-9474(99)00844-1
  [hep-ph/9805306].

\bibitem{Jido:2001nt} 
  D.~Jido, M.~Oka and A.~Hosaka,
  ``Chiral symmetry of baryons,''
  Prog.\ Theor.\ Phys.\  {\bf 106}, 873 (2001)
  doi:10.1143/PTP.106.873
  [hep-ph/0110005].

\bibitem{Gallas:2009qp} 
  S.~Gallas, F.~Giacosa and D.~H.~Rischke,
  ``Vacuum phenomenology of the chiral partner of the nucleon in a linear sigma model with vector mesons,''
  Phys.\ Rev.\ D {\bf 82}, 014004 (2010)
  doi:10.1103/PhysRevD.82.014004
  [arXiv:0907.5084 [hep-ph]].



\bibitem{Nemoto:1998um} 
  Y.~Nemoto, D.~Jido, M.~Oka and A.~Hosaka,
  ``Decays of 1/2- baryons in chiral effective theory,''
  Phys.\ Rev.\ D {\bf 57}, 4124 (1998)
  doi:10.1103/PhysRevD.57.4124
  [hep-ph/9710445].

\bibitem{Chen:2008qv} 
  H.~X.~Chen, V.~Dmitrasinovic, A.~Hosaka, K.~Nagata and S.~L.~Zhu,
  ``Chiral Properties of Baryon Fields with Flavor SU(3) Symmetry,''
  Phys.\ Rev.\ D {\bf 78}, 054021 (2008)
  doi:10.1103/PhysRevD.78.054021
  [arXiv:0806.1997 [hep-ph]].

\bibitem{Dmitrasinovic:2009vp} 
  V.~Dmitrasinovic, A.~Hosaka and K.~Nagata,
  ``Nucleon axial couplings and [(1/2,0) + (0,1/2)] - [(1,1/2) + (1/2,1)] chiral multiplet mixing,''
  Mod.\ Phys.\ Lett.\ A {\bf 25}, 233 (2010)
  doi:10.1142/S0217732310032494
  [arXiv:0912.2372 [hep-ph]].

\bibitem{Dmitrasinovic:2009vy} 
  V.~Dmitrasinovic, A.~Hosaka and K.~Nagata,
  ``A Lagrangian for the Chiral (1/2,0) + (0,1/2) Quartet Nucleon Resonances,''
  Int.\ J.\ Mod.\ Phys.\ E {\bf 19}, 91 (2010)
  doi:10.1142/S0218301310014650
  [arXiv:0912.2396 [hep-ph]].

\bibitem{Chen:2009sf} 
  H.~X.~Chen, V.~Dmitrasinovic and A.~Hosaka,
  ``Baryon fields with U(L)(3) X U(R)(3) chiral symmetry II: Axial currents of nucleons and hyperons,''
  Phys.\ Rev.\ D {\bf 81}, 054002 (2010)
  doi:10.1103/PhysRevD.81.054002
  [arXiv:0912.4338 [hep-ph]].

\bibitem{Chen:2010ba} 
  H.~X.~Chen, V.~Dmitrasinovic and A.~Hosaka,
  ``Baryon Fields with $U_L(3) times U_R(3)$ Chiral Symmetry III: Interactions with Chiral $(3,\bar{3})+ (\bar{3},3)$ Spinless Mesons,''
  Phys.\ Rev.\ D {\bf 83}, 014015 (2011)
  doi:10.1103/PhysRevD.83.014015
  [arXiv:1009.2422 [hep-ph]].

\bibitem{Steinheimer:2011ea} 
  J.~Steinheimer, S.~Schramm and H.~Stocker,
  ``The hadronic SU(3) Parity Doublet Model for Dense Matter, its extension to quarks and the strange equation of state,''
  Phys.\ Rev.\ C {\bf 84}, 045208 (2011)
  doi:10.1103/PhysRevC.84.045208
  [arXiv:1108.2596 [hep-ph]].

\bibitem{Chen:2011rh} 
  H.~X.~Chen, V.~Dmitrasinovic and A.~Hosaka,
  ``Baryons with $\mbox{U}_L(3)\times \mbox{U}_R(3)$ Chiral Symmetry IV: Interactions with Chiral $(8,1)+(1,8)$ Vector and Axial-vector Mesons and Anomalous Magnetic Moments,''
  Phys.\ Rev.\ C {\bf 85}, 055205 (2012)
  doi:10.1103/PhysRevC.85.055205
  [arXiv:1109.3130 [hep-ph]].

\bibitem{Nishihara:2015fka} 
  H.~Nishihara and M.~Harada,
  ``Extended Goldberger-Treiman relation in a three-flavor parity doublet model,''
  Phys.\ Rev.\ D {\bf 92}, no. 5, 054022 (2015)
  doi:10.1103/PhysRevD.92.054022
  [arXiv:1506.07956 [hep-ph]].

\bibitem{Olbrich:2015gln} 
  L.~Olbrich, M.~Z\`et\'enyi, F.~Giacosa and D.~H.~Rischke,
  ``Three-flavor chiral effective model with four baryonic multiplets within the mirror assignment,''
  Phys.\ Rev.\ D {\bf 93}, no. 3, 034021 (2016)
  doi:10.1103/PhysRevD.93.034021
  [arXiv:1511.05035 [hep-ph]].

\bibitem{Dmitrasinovic:2016hup} 
  V.~Dmitra\v{s}inovi\'c, H.~X.~Chen and A.~Hosaka,
  ``Baryon fields with UL(3)×UR(3) chiral symmetry. V. Pion-nucleon and kaon-nucleon Σ terms,''
  Phys.\ Rev.\ C {\bf 93}, no. 6, 065208 (2016)
  doi:10.1103/PhysRevC.93.065208
  [arXiv:1812.03414 [hep-ph]].

\bibitem{Sasaki:2017glk} 
  C.~Sasaki,
  ``Parity doubling of baryons in a chiral approach with three flavors,''
  Nucl.\ Phys.\ A {\bf 970}, 388 (2018)
  doi:10.1016/j.nuclphysa.2018.01.004
  [arXiv:1707.05081 [hep-ph]].

\bibitem{Yamazaki:2018stk} 
  T.~Yamazaki and M.~Harada,
  ``Chiral partner structure of light nucleons in an extended parity doublet model,''
  arXiv:1809.02359 [hep-ph].



\bibitem{Hatsuda:1988mv} 
  T.~Hatsuda and M.~Prakash,
  ``Parity Doubling of the Nucleon and First Order Chiral Transition in Dense Matter,''
  Phys.\ Lett.\ B {\bf 224}, 11 (1989).
  doi:10.1016/0370-2693(89)91040-X

\bibitem{Zschiesche:2006zj} 
  D.~Zschiesche, L.~Tolos, J.~Schaffner-Bielich and R.~D.~Pisarski,
  ``Cold, dense nuclear matter in a SU(2) parity doublet model,''
  Phys.\ Rev.\ C {\bf 75}, 055202 (2007)
  doi:10.1103/PhysRevC.75.055202
  [nucl-th/0608044].

\bibitem{Dexheimer:2007tn} 
  V.~Dexheimer, S.~Schramm and D.~Zschiesche,
  ``Nuclear matter and neutron stars in a parity doublet model,''
  Phys.\ Rev.\ C {\bf 77}, 025803 (2008)
  doi:10.1103/PhysRevC.77.025803
  [arXiv:0710.4192 [nucl-th]].

\bibitem{Dexheimer:2008cv} 
  V.~Dexheimer, G.~Pagliara, L.~Tolos, J.~Schaffner-Bielich and S.~Schramm,
  ``Neutron stars within the SU(2) parity doublet model,''
  Eur.\ Phys.\ J.\ A {\bf 38}, 105 (2008)
  doi:10.1140/epja/i2008-10652-0
  [arXiv:0805.3301 [nucl-th]].

\bibitem{Sasaki:2010bp} 
  C.~Sasaki and I.~Mishustin,
  ``Thermodynamics of dense hadronic matter in a parity doublet model,''
  Phys.\ Rev.\ C {\bf 82}, 035204 (2010)
  doi:10.1103/PhysRevC.82.035204
  [arXiv:1005.4811 [hep-ph]].

\bibitem{Sasaki:2011ff} 
  C.~Sasaki, H.~K.~Lee, W.~G.~Paeng and M.~Rho,
  ``Conformal anomaly and the vector coupling in dense matter,''
  Phys.\ Rev.\ D {\bf 84}, 034011 (2011)
  doi:10.1103/PhysRevD.84.034011
  [arXiv:1103.0184 [hep-ph]].

\bibitem{Gallas:2011qp} 
  S.~Gallas, F.~Giacosa and G.~Pagliara,
  ``Nuclear matter within a dilatation-invariant parity doublet model: the role of the tetraquark at nonzero density,''
  Nucl.\ Phys.\ A {\bf 872}, 13 (2011)
  doi:10.1016/j.nuclphysa.2011.09.008
  [arXiv:1105.5003 [hep-ph]].

\bibitem{Paeng:2011hy} 
  W.~G.~Paeng, H.~K.~Lee, M.~Rho and C.~Sasaki,
  ``Dilaton-Limit Fixed Point in Hidden Local Symmetric Parity Doublet Model,''
  Phys.\ Rev.\ D {\bf 85}, 054022 (2012)
  doi:10.1103/PhysRevD.85.054022
  [arXiv:1109.5431 [hep-ph]].

\bibitem{Dexheimer:2012eu} 
  V.~Dexheimer, J.~Steinheimer, R.~Negreiros and S.~Schramm,
  ``Hybrid Stars in an SU(3) parity doublet model,''
  Phys.\ Rev.\ C {\bf 87}, no. 1, 015804 (2013)
  doi:10.1103/PhysRevC.87.015804
  [arXiv:1206.3086 [astro-ph.HE]].

\bibitem{Paeng:2013xya} 
  W.~G.~Paeng, H.~K.~Lee, M.~Rho and C.~Sasaki,
  ``Interplay between $\omega$-nucleon interaction and nucleon mass in dense baryonic matter,''
  Phys.\ Rev.\ D {\bf 88}, 105019 (2013)
  doi:10.1103/PhysRevD.88.105019
  [arXiv:1303.2898 [nucl-th]].

\bibitem{Benic:2015pia} 
  S.~Benic, I.~Mishustin and C.~Sasaki,
  ``Effective model for the QCD phase transitions at finite baryon density,''
  Phys.\ Rev.\ D {\bf 91}, no. 12, 125034 (2015)
  doi:10.1103/PhysRevD.91.125034
  [arXiv:1502.05969 [hep-ph]].

\bibitem{Motohiro:2015taa} 
  Y.~Motohiro, Y.~Kim and M.~Harada,
  ``Asymmetric nuclear matter in a parity doublet model with hidden local symmetry,''
  Phys.\ Rev.\ C {\bf 92}, no. 2, 025201 (2015)
  Erratum: [Phys.\ Rev.\ C {\bf 95}, no. 5, 059903 (2017)]
  doi:10.1103/PhysRevC.92.025201, 10.1103/PhysRevC.95.059903
  [arXiv:1505.00988 [nucl-th]].


\bibitem{Mukherjee:2016nhb} 
  A.~Mukherjee, J.~Steinheimer and S.~Schramm,
  ``Higher-order baryon number susceptibilities: interplay between the chiral and the nuclear liquid-gas transitions,''
  Phys.\ Rev.\ C {\bf 96}, no. 2, 025205 (2017)
  doi:10.1103/PhysRevC.96.025205
  [arXiv:1611.10144 [nucl-th]].

\bibitem{Suenaga:2017wbb} 
  D.~Suenaga,
  ``Examination of $N^*(1535)$ as a probe to observe the partial restoration of chiral symmetry in nuclear matter,''
  Phys.\ Rev.\ C {\bf 97}, no. 4, 045203 (2018)
  doi:10.1103/PhysRevC.97.045203
  [arXiv:1704.03630 [nucl-th]].

\bibitem{Takeda:2017mrm} 
  Y.~Takeda, Y.~Kim and M.~Harada,
  ``Catalysis of partial chiral symmetry restoration by $\Delta$ matter,''
  Phys.\ Rev.\ C {\bf 97}, no. 6, 065202 (2018)
  doi:10.1103/PhysRevC.97.065202
  [arXiv:1704.04357 [nucl-th]].

\bibitem{Mukherjee:2017jzi} 
  A.~Mukherjee, S.~Schramm, J.~Steinheimer and V.~Dexheimer,
  ``The application of the Quark-Hadron Chiral Parity-Doublet Model to neutron star matter,''
  Astron.\ Astrophys.\  {\bf 608}, A110 (2017)
  doi:10.1051/0004-6361/201731505
  [arXiv:1706.09191 [nucl-th]].

\bibitem{Marczenko:2017huu} 
  M.~Marczenko and C.~Sasaki,
  ``Net-baryon number fluctuations in the Hybrid Quark-Meson-Nucleon model at finite density,''
  Phys.\ Rev.\ D {\bf 97}, no. 3, 036011 (2018)
  doi:10.1103/PhysRevD.97.036011
  [arXiv:1711.05521 [hep-ph]].



\bibitem{Abuki:2018ijb} 
  H.~Abuki, Y.~Takeda and M.~Harada,
  ``Dual chiral density waves in nuclear matter,''
  EPJ Web Conf.\  {\bf 192}, 00020 (2018)
  doi:10.1051/epjconf/201819200020
  [arXiv:1809.06485 [hep-ph]].


\bibitem{Marczenko:2018jui} 
  M.~L.~Marczenko, D.~Blaschke, K.~Redlich and C.~Sasaki,
  ``Chiral symmetry restoration by parity doubling and the structure of neutron stars,''
  Phys.\ Rev.\ D {\bf 98}, no. 10, 103021 (2018)
  doi:10.1103/PhysRevD.98.103021
  [arXiv:1805.06886 [nucl-th]].

\bibitem{TheLIGOScientific:2017qsa} 
  B.~P.~Abbott {\it et al.} [LIGO Scientific and Virgo Collaborations],
  ``GW170817: Observation of Gravitational Waves from a Binary Neutron Star Inspiral,''
  Phys.\ Rev.\ Lett.\  {\bf 119}, no. 16, 161101 (2017)
  doi:10.1103/PhysRevLett.119.161101
  [arXiv:1710.05832 [gr-qc]].

\bibitem{GBM:2017lvd} 
  B.~P.~Abbott {\it et al.} [LIGO Scientific and Virgo and Fermi GBM and INTEGRAL and IceCube and IPN and Insight-Hxmt and ANTARES and Swift and Dark Energy Camera GW-EM and DES and DLT40 and GRAWITA and Fermi-LAT and ATCA and ASKAP and OzGrav and DWF (Deeper Wider Faster Program) and AST3 and CAASTRO and VINROUGE and MASTER and J-GEM and GROWTH and JAGWAR and CaltechNRAO and TTU-NRAO and NuSTAR and Pan-STARRS and KU and Nordic Optical Telescope and ePESSTO and GROND and Texas Tech University and TOROS and BOOTES and MWA and CALET and IKI-GW Follow-up and H.E.S.S. and LOFAR and LWA and HAWC and Pierre Auger and ALMA and Pi of Sky and DFN and ATLAS Telescopes and High Time Resolution Universe Survey and RIMAS and RATIR and SKA South Africa/MeerKAT Collaborations and AstroSat Cadmium Zinc Telluride Imager Team and AGILE Team and 1M2H Team and Las Cumbres Observatory Group and MAXI Team and TZAC Consortium and SALT Group and Euro VLBI Team and Chandra Team at McGill University],
  ``Multi-messenger Observations of a Binary Neutron Star Merger,''
  Astrophys.\ J.\  {\bf 848}, no. 2, L12 (2017)
  doi:10.3847/2041-8213/aa91c9
  [arXiv:1710.05833 [astro-ph.HE]].

\bibitem{Abbott:2018exr} 
  B.~P.~Abbott {\it et al.} [LIGO Scientific and Virgo Collaborations],
  ``GW170817: Measurements of neutron star radii and equation of state,''
  Phys.\ Rev.\ Lett.\  {\bf 121}, no. 16, 161101 (2018)
  doi:10.1103/PhysRevLett.121.161101
  [arXiv:1805.11581 [gr-qc]].




\bibitem{Shin:2018axs} 
  I.~J.~Shin, W.~G.~Paeng, M.~Harada and Y.~Kim,
  ``Nuclear structure in Parity Doublet Model,''
  arXiv:1805.03402 [nucl-th].




\bibitem{Aarts:2017rrl} 
  G.~Aarts, C.~Allton, D.~De Boni, S.~Hands, B.~J\"ager, C.~Praki and J.~I.~Skullerud,
  ``Light baryons below and above the deconfinement transition: medium effects and parity doubling,''
  JHEP {\bf 1706}, 034 (2017)
  doi:10.1007/JHEP06(2017)034
  [arXiv:1703.09246 [hep-lat]].

\bibitem{Aarts:2017iai} 
  G.~Aarts, C.~Allton, D.~de Boni, S.~Hands, B.~J\"ager, C.~Praki and J.~I.~Skullerud,
  ``Baryons in the plasma: in-medium effects and parity doubling,''
  EPJ Web Conf.\  {\bf 171}, 14005 (2018)
  doi:10.1051/epjconf/201817114005
  [arXiv:1710.00566 [hep-lat]].

\bibitem{Aarts:2018glk} 
  G.~Aarts, C.~Allton, D.~De Boni and B.~J\"ager,
  ``Hyperons in thermal QCD: a lattice view,''
  arXiv:1812.07393 [hep-lat].




\bibitem{Bando:1987br} 
  M.~Bando, T.~Kugo and K.~Yamawaki,
  ``Nonlinear Realization and Hidden Local Symmetries,''
  Phys.\ Rept.\  {\bf 164}, 217 (1988).
  doi:10.1016/0370-1573(88)90019-1

\bibitem{Harada:2003jx} 
  M.~Harada and K.~Yamawaki,
  ``Hidden local symmetry at loop: A New perspective of composite gauge boson and chiral phase transition,''
  Phys.\ Rept.\  {\bf 381}, 1 (2003)
  doi:10.1016/S0370-1573(03)00139-X
  [hep-ph/0302103].


\bibitem{Tolman:1939jz} 
  R.~C.~Tolman,
  ``Static solutions of Einstein's field equations for spheres of fluid,''
  Phys.\ Rev.\  {\bf 55}, 364 (1939).
  doi:10.1103/PhysRev.55.364

\bibitem{Oppenheimer:1939ne} 
  J.~R.~Oppenheimer and G.~M.~Volkoff,
  ``On Massive neutron cores,''
  Phys.\ Rev.\  {\bf 55}, 374 (1939).
  doi:10.1103/PhysRev.55.374
	

\bibitem{Hinderer:2007mb} 
  T.~Hinderer,
  ``Tidal Love numbers of neutron stars,''
  Astrophys.\ J.\  {\bf 677}, 1216 (2008)
  doi:10.1086/533487
  [arXiv:0711.2420 [astro-ph]].

\bibitem{Hinderer:2009ca} 
  T.~Hinderer, B.~D.~Lackey, R.~N.~Lang and J.~S.~Read,
  ``Tidal deformability of neutron stars with realistic equations of state and their gravitational wave signatures in binary inspiral,''
  Phys.\ Rev.\ D {\bf 81}, 123016 (2010)
  doi:10.1103/PhysRevD.81.123016
  [arXiv:0911.3535 [astro-ph.HE]].

\bibitem{Malik:2018zcf} 
  T.~Malik, N.~Alam, M.~Fortin, C.~Provid\^encia, B.~K.~Agrawal, T.~K.~Jha, B.~Kumar and S.~K.~Patra,
  ``GW170817: constraining the nuclear matter equation of state from the neutron star tidal deformability,''
  Phys.\ Rev.\ C {\bf 98}, no. 3, 035804 (2018)
  doi:10.1103/PhysRevC.98.035804
  [arXiv:1805.11963 [nucl-th]].


\bibitem{Takahashi:2008fy} 
  T.~T.~Takahashi and T.~Kunihiro,
  ``Axial charges of N(1535) and N(1650) in lattice QCD with two flavors of dynamical quarks,''
  Phys.\ Rev.\ D {\bf 78}, 011503 (2008)
  doi:10.1103/PhysRevD.78.011503
  [arXiv:0801.4707 [hep-lat]].



\bibitem{Baym:1971pw} 
  G.~Baym, C.~Pethick and P.~Sutherland,
  ``The Ground state of matter at high densities: Equation of state and stellar models,''
  Astrophys.\ J.\  {\bf 170}, 299 (1971).
  doi:10.1086/151216

\bibitem{Sharma:2015bna} 
  B.~K.~Sharma, M.~Centelles, X.~Vi\~nas, M.~Baldo and G.~F.~Burgio,
  ``Unified equation of state for neutron stars on a microscopic basis,''
  Astron.\ Astrophys.\  {\bf 584}, A103 (2015)
  doi:10.1051/0004-6361/201526642
  [arXiv:1506.00375 [nucl-th]].

\bibitem{Abbott:2018wiz} 
  B.~P.~Abbott {\it et al.} [LIGO Scientific and Virgo Collaborations],
  ``Properties of the binary neutron star merger GW170817,''
  Phys.\ Rev.\ X {\bf 9}, no. 1, 011001 (2019)
  doi:10.1103/PhysRevX.9.011001
  [arXiv:1805.11579 [gr-qc]].

\bibitem{Demorest:2010bx} 
  P.~Demorest, T.~Pennucci, S.~Ransom, M.~Roberts and J.~Hessels,
  ``Shapiro Delay Measurement of A Two Solar Mass Neutron Star,''
  Nature {\bf 467}, 1081 (2010)
  doi:10.1038/nature09466
  [arXiv:1010.5788 [astro-ph.HE]].

\bibitem{Antoniadis:2013pzd} 
  J.~Antoniadis {\it et al.},
  ``A Massive Pulsar in a Compact Relativistic Binary,''
  Science {\bf 340}, 6131 (2013)
  doi:10.1126/science.1233232
  [arXiv:1304.6875 [astro-ph.HE]].


\bibitem{Masuda:2012kf} 
  K.~Masuda, T.~Hatsuda and T.~Takatsuka,
  ``Hadron-Quark Crossover and Massive Hybrid Stars with Strangeness,''
  Astrophys.\ J.\  {\bf 764}, 12 (2013)
  doi:10.1088/0004-637X/764/1/12
  [arXiv:1205.3621 [nucl-th]].

\bibitem{Masuda:2012ed} 
  K.~Masuda, T.~Hatsuda and T.~Takatsuka,
  ``Hadron-quark crossover and massive hybrid stars,''
  PTEP {\bf 2013}, no. 7, 073D01 (2013)
  doi:10.1093/ptep/ptt045
  [arXiv:1212.6803 [nucl-th]].

\bibitem{Masuda:2015wva} 
  K.~Masuda, T.~Hatsuda and T.~Takatsuka,
  ``Hadron-quark crossover and hot neutron stars at birth,''
  PTEP {\bf 2016}, no. 2, 021D01 (2016)
  doi:10.1093/ptep/ptv187
  [arXiv:1506.00984 [nucl-th]].

\bibitem{Masuda:2015kha} 
  K.~Masuda, T.~Hatsuda and T.~Takatsuka,
  ``Hyperon Puzzle, Hadron-Quark Crossover and Massive Neutron Stars,''
  Eur.\ Phys.\ J.\ A {\bf 52}, no. 3, 65 (2016)
  doi:10.1140/epja/i2016-16065-6
  [arXiv:1508.04861 [nucl-th]].

\bibitem{Kojo:2014rca} 
  T.~Kojo, P.~D.~Powell, Y.~Song and G.~Baym,
  ``Phenomenological QCD equation of state for massive neutron stars,''
  Phys.\ Rev.\ D {\bf 91}, no. 4, 045003 (2015)
  doi:10.1103/PhysRevD.91.045003
  [arXiv:1412.1108 [hep-ph]].

\bibitem{Kojo:2015nzn} 
  T.~Kojo, P.~D.~Powell, Y.~Song and G.~Baym,
  ``Phenomenological QCD equations of state for neutron stars,''
  Nucl.\ Phys.\ A {\bf 956}, 821 (2016)
  doi:10.1016/j.nuclphysa.2016.02.008
  [arXiv:1512.08592 [hep-ph]].

\bibitem{Baym:2017whm} 
  G.~Baym, T.~Hatsuda, T.~Kojo, P.~D.~Powell, Y.~Song and T.~Takatsuka,
  ``From hadrons to quarks in neutron stars: a review,''
  Rept.\ Prog.\ Phys.\  {\bf 81}, no. 5, 056902 (2018)
  doi:10.1088/1361-6633/aaae14
  [arXiv:1707.04966 [astro-ph.HE]].

\bibitem{Wu:2018kww} 
  X.~Wu, A.~Ohnishi and H.~Shen,
  ``Effects of quark-matter symmetry energy on hadron-quark coexistence in neutron-star matter,''
  Phys.\ Rev.\ C {\bf 98}, no. 6, 065801 (2018)
  doi:10.1103/PhysRevC.98.065801
  [arXiv:1806.03760 [nucl-th]].



\bibitem{Kolomeitsev:2016sjl} 
  I.~Tews, J.~M.~Lattimer, A.~Ohnishi and E.~E.~Kolomeitsev,
  ``Symmetry Parameter Constraints from a Lower Bound on Neutron-matter Energy,''
  Astrophys.\ J.\  {\bf 848}, no. 2, 105 (2017)
  doi:10.3847/1538-4357/aa8db9
  [arXiv:1611.07133 [nucl-th]].

\bibitem{Ohnishi:private}
A.~Ohnishi, private communication.

\bibitem{Most:2018hfd} 
  E.~R.~Most, L.~R.~Weih, L.~Rezzolla and J.~Schaffner-Bielich,
  ``New constraints on radii and tidal deformabilities of neutron stars from GW170817,''
  Phys.\ Rev.\ Lett.\  {\bf 120}, no. 26, 261103 (2018)
  doi:10.1103/PhysRevLett.120.261103
  [arXiv:1803.00549 [gr-qc]].

\bibitem{Fattoyev:2017jql} 
  F.~J.~Fattoyev, J.~Piekarewicz and C.~J.~Horowitz,
  ``Neutron Skins and Neutron Stars in the Multimessenger Era,''
  Phys.\ Rev.\ Lett.\  {\bf 120}, no. 17, 172702 (2018)
  doi:10.1103/PhysRevLett.120.172702
  [arXiv:1711.06615 [nucl-th]].

\bibitem{Kim:2018aoi} 
  Y.~M.~Kim, Y.~Lim, K.~Kwak, C.~H.~Hyun and C.~H.~Lee,
  ``Tidal Deformability of Neutron Stars with Realistic Nuclear Energy Density Functionals,''
  Phys.\ Rev.\ C {\bf 98}, no. 6, 065805 (2018)
  doi:10.1103/PhysRevC.98.065805
  [arXiv:1805.00219 [nucl-th]].


\bibitem{Steiner:2010fz} 
  A.~W.~Steiner, J.~M.~Lattimer and E.~F.~Brown,
  ``The Equation of State from Observed Masses and Radii of Neutron Stars,''
  Astrophys.\ J.\  {\bf 722}, 33 (2010)
  doi:10.1088/0004-637X/722/1/33
  [arXiv:1005.0811 [astro-ph.HE]].


\bibitem{Lattimer:2014sga} 
  J.~M.~Lattimer and A.~W.~Steiner,
  ``Constraints on the symmetry energy using the mass-radius relation of neutron stars,''
  Eur.\ Phys.\ J.\ A {\bf 50}, 40 (2014)
  doi:10.1140/epja/i2014-14040-y
  [arXiv:1403.1186 [nucl-th]].

\bibitem{Togashi:2017mjp} 
  H.~Togashi, K.~Nakazato, Y.~Takehara, S.~Yamamuro, H.~Suzuki and M.~Takano,
  Nucl.\ Phys.\ A {\bf 961}, 78 (2017)
  doi:10.1016/j.nuclphysa.2017.02.010
  [arXiv:1702.05324 [nucl-th]].

\bibitem{Ma:2016nki} 
  Y.~L.~Ma and M.~Rho,
  ``Scale-chiral symmetry, $\omega$ meson and dense baryonic matter,''
  Phys.\ Rev.\ D {\bf 97}, no. 9, 094017 (2018)
  doi:10.1103/PhysRevD.97.094017
  [arXiv:1612.04079 [nucl-th]].

\bibitem{Harada:2016uca} 
  M.~Harada, Y.~L.~Ma, D.~Suenaga and Y.~Takeda,
  ``Relation between the mass modification of the heavy-light mesons and the chiral symmetry structure in dense matter,''
  Progress of Theoretical and Experimental Physics, Volume 2017,
  Issue 11, 1 November 2017, 113D01
  doi:10.1093/ptep/ptx140
  [arXiv:1612.03496 [hep-ph]].

\bibitem{Suenaga:2018kta} 
  D.~Suenaga,
  ``Spectral function for $\bar{D}_{0}^{\ast}$ $(0^+)$ meson in isospin asymmetric nuclear matter with chiral partner structure,''
  arXiv:1805.01709 [nucl-th].


\end{thebibliography}
\end{document}